\documentclass[nofootinbib,preprint,tightenlines,superscriptaddress]{revtex4}

\usepackage[utf8]{inputenc}
\usepackage{amssymb,amsthm,amsmath,amstext,amsbsy,bbm}
\usepackage{url}
\usepackage{nicefrac}
\usepackage{graphicx}
\usepackage{ifpdf}

\newcommand{\ie}{\textit{i.e.}}

\newcommand{\cf}{\textit{cf.}}

\newcommand{\mathspace}{\ \ }
\newcommand{\mathtext}[1]{\mathspace\text{#1}\mathspace}

\newcommand{\bra}[1]{\left\langle #1\right|}
\newcommand{\ket}[1]{\left|#1\right\rangle}
\newcommand{\braket}[2]{\left\langle #1\middle|#2\right\rangle}

\newcommand{\mbraketbig}[3]{\big\langle #1\big|#2\big|#3\big\rangle}

\newcommand{\dd}{\mathrm{d}}
\newcommand{\ii}{\mathrm{i}}
\newcommand{\ee}{\mathrm{e}}

\newcommand{\ZZ}{\mathbb{Z}}

\newcommand{\OO}{\mathcal{O}}
\newcommand{\YY}{Y}

\newcommand{\vdelta}{\delta^{(3)}}

\newcommand{\Laplace}{\Delta}

\newcommand{\vecr}{\mathbf{r}}
\newcommand{\vecx}{\mathbf{x}}

\newcommand{\vecn}{\mathbf{n}}
\newcommand{\vecm}{\mathbf{m}}
\newcommand{\vecp}{\mathbf{p}}

\newcommand{\vecq}{\mathbf{q}}
\newcommand{\vece}{\mathbf{e}}

\newcommand{\vZero}{\mathbf{0}}
\newcommand{\vNabla}{\boldsymbol{\nabla}}

\newcommand{\ddd}[1]{\frac{\dd}{\dd #1}}
\newcommand{\del}[1]{\frac{\partial}{\partial #1}}

\newcommand{\ddr}{\mathrm{d}^3r}

\begin{document}

\title{Non-relativistic bound states in a finite volume}

\author{Sebastian K\"onig}
\email{koenig@hiskp.uni-bonn.de}
\affiliation{Helmholtz-Institut f\"ur Strahlen- und Kernphysik (Theorie)\\
and Bethe Center for Theoretical Physics, Universit\"at Bonn, 53115 Bonn,
Germany\\[0.5em]}

\author{Dean Lee}
\email{dean_lee@ncsu.edu}
\affiliation{Department of Physics, North Carolina State University, Raleigh,
NC 27695, USA}
\affiliation{Helmholtz-Institut f\"ur Strahlen- und Kernphysik (Theorie)\\
and Bethe Center for Theoretical Physics, Universit\"at Bonn, 53115 Bonn,
Germany\\[0.5em]}

\author{H.-W. Hammer}
\email{hammer@hiskp.uni-bonn.de}
\affiliation{Helmholtz-Institut f\"ur Strahlen- und Kernphysik (Theorie)\\
and Bethe Center for Theoretical Physics, Universit\"at Bonn, 53115 Bonn,
Germany\\[0.5em]}

\date{\today}

\begin{abstract}
We derive general results for the mass shift of bound states with angular 
momentum $\ell \geq 1$ in a periodic cubic box in two and three spatial
dimensions.  Our results have applications to lattice simulations of hadronic
molecules, halo nuclei, and Feshbach molecules.  The sign of the mass shift can
be related to the symmetry properties of the state under consideration.  We
verify our analytical results with explicit numerical calculations.  Moreover,
we comment on the relations connecting  the effective range parameter, the
binding momentum of a given state and the asymptotic normalization coefficient
of the corresponding wave function.  We give explicit expressions for this
relation in the shallow binding limit.
\end{abstract}

\maketitle

\section{Introduction}
\label{sec:Intro}

Lattice simulations are used in many areas of quantum physics, ranging from
nuclear and particle physics to atomic and condensed matter
physics~\cite{Lee:2008fa,Bazavov:2009bb,Beane:2010em}.  In such calculations the
system is solved numerically using a discrete spacetime lattice over a finite
volume.  In practice this finite volume is usually taken to be a cubic box with 
periodic boundaries.  When simulating composite objects such as bound states,
the boundaries of the periodic box will modify quantum wave functions.  This 
leads to finite volume shifts in the binding energies, and detailed knowledge
of these finite volume effects is necessary to improve high precision lattice
calculations.

In Ref.~\cite{Luscher:1985dn}, L\"uscher derived a formula for the 
finite volume mass shift of S-wave bound states of two particles with reduced
mass $\mu$ interacting via a potential with finite range $R$.  When a bound
state with energy $-{E_B}$ is put in a periodic cubic box of length $L$, its
energy in the rest frame is shifted by an amount
\begin{equation}
 \Delta m_B = -3|\gamma|^2\frac{\ee^{-{\kappa L}}}{\mu L}
 + \OO\big(\ee^{-{\sqrt2\kappa L}}\big) \,,
\label{eq:Delta_m-Luscher}
\end{equation}
where $\displaystyle\kappa=\sqrt{2\mu E_B}$ is the binding momentum and 
$\gamma$ is the asymptotic wave function normalization defined by $\psi_B(r) =
\gamma\,\ee^{-{\kappa r}}/(\sqrt{4 \pi}r)$ for $r>R$.  For potentials with
exponential fall-off, $V(r)\sim\exp(-r/R)$ for large $r$, the formula is
modified by exponentially small corrections when the binding momentum $\kappa$
is smaller than $1/R$.  Bound states moving in a finite periodic volume also
have a topological phase correction to the
energy~\cite{Bour:2011ef,Davoudi:2011md}.  This topological phase correction
contains information about the number and mass of the constituents of the bound
states.  It must be included when determining scattering phase shifts for
composite objects in a finite volume.

In Ref.~\cite{Konig:2011nz}, we briefly discussed the generalization of
L\"uscher's formula (\ref{eq:Delta_m-Luscher}) for the finite volume mass shift
to higher partial waves.  In this paper, we present the full derivation
of these results.  Moreover, we give the general mass shift for states with
angular momenta up to $\ell=3$.  In general, the mass shift for a given state
depends on its transformation properties  with respect to the symmetry group of
the cubic box.  In addition to reducing finite volume effects from precision
lattice calculations, our finite volume results can also be used as a diagnostic
tool to probe the angular momentum and radial structure of the bound state wave
function.  A summary of our results is given by Eq.~\eqref{eq:Delta_m-Gamma}
together with Table~\ref{tab:Results} at the end of Sec.~\ref{sec:Results}.  We
also derive a general mass shift formula for two-dimensional systems in a finite
area with periodic boundaries.

In our discussion here we only consider stable bound states.  However,
resonances above continuum thresholds can also be described from the
volume dependence of avoided level
crossings~\cite{Luscher:1991cf,Bernard:2008ax}.  Our results are universal and
can be applied to a wide range of systems. We will discuss a few examples
below.

In particle physics, there is some interest in hadronic molecules with 
angular momentum~\cite{Novikov:1977dq,Bulava:2009jb,Matsuyama:2006rp}.  
In the case of S-waves, the deuteron and some exotic weakly bound states
such as the H-dibaryon were recently studied in lattice QCD~\cite{Beane:2011iw}.
Similar investigations for exotic bound states with angular momentum appear
feasible in the future.  In atomic physics, several experiments have
investigated strongly-interacting P-wave Feshbach resonances in $^{6}$Li and
$^{40}$K~\cite{Regal:2003zz,Schunk:2005A,Gaebler:2007A}, which can be tuned to
produce bound P-wave dimers.  If such systems are simulated in a finite volume,
our results can be used to describe the finite volume dependence of the dimers.

Halo nuclei constitute a special class of weakly-bound nuclei with molecular
character.  They consist of a compact core and one or more nucleons with low
separation energy~\cite{Riisager-94}.  Among halo nuclei there are some systems
with nonzero orbital angular momentum.  A well-known example of a P-wave halo
state is the $J^P=1/2^-$ excited state in $^{11}$Be.  The electromagnetic
properties of the low-lying states in this nucleus can be well described in a
two-body halo picture of a $^{10}$Be core and a
neutron~\cite{Typel:2004zm,Hammer:2011ye}.  A similar strategy was applied
to radiative neutron capture on $^7$Li~\cite{Rupak:2011nk}.  A related class of
systems is given by nuclei with an $\alpha$-cluster structure such as $^8$Be and
excited states of $^{12}$C~\cite{Tohsaki:2001,Chernykh:2007,Epelbaum:2011md}.
The Hoyle state in $^{12}$C was recently calculated for the first time in an
{\it ab initio} approach using chiral effective field theory and nuclear lattice
calculations~\cite{Epelbaum:2011md}.  Understanding the finite volume
corrections is crucial in such a calculation.  Due to the $\alpha$-cluster
structure, the volume dependence of the Hoyle state is governed by the volume
dependence of a three-body system.  Finally, we note that the asymptotic
normalization coefficient of the bound state wave function appears in the mass
shift formula. Hence, our results can be used to extract this quantity from
lattice calculations at finite volume.  The asymptotic normalization is directly
connected to zero-energy capture reactions~\cite{Xu:1994zz}, which play an
important role in nuclear astrophysics.  

The paper is organized as follows. We start with some prerequisites and a
general discussion of the finite volume mass shift in
Section~\ref{sec:BoundStates}.  L\"uscher's result for S-waves is reviewed in 
Section \ref{sec:S-wave}, while our extension to higher partial waves is given
in Section \ref{sec:P-wave-and-higher}.  In particular, we discuss the mass
shift for the irreducible representations of the cubic group, relate the sign of
the shift to the leading parity, and derive a trace formula for the
multiplet-averaged mass shift for states with arbitrary angular momentum $\ell$.
In Section \ref{sec:NumTest}, we verify our results numerically for two model
systems.  The case of two spatial dimensions is treated in Section~\ref{sec:2D}.
Finally, a brief summary and outlook is given in Section \ref{sec:Conclusion}.
In the appendix, we discuss the relation of the asymptotic normalization
constant to scattering parameters and derive explicit shallow binding relations
between the effective range, the binding momentum and the asymptotic
normalization coefficient.

\section{Bound states in a finite volume}
\label{sec:BoundStates}

As a starting point, we first review several results from~\cite{Luscher:1985dn}.  
We closely follow L\"uscher's derivation, but consider a more general system
with arbitrary angular momentum and non-local interactions.

\subsection{Definitions and basic identities}
\label{sec:Prereq}

We start with some definitions and basic identities about bound states in
rotationally symmetric potentials $V(r)$.  This will also help to define our
notation used throughout the discussion.  For a bound state with energy $E=-E_B$
and angular-momentum quantum numbers $(\ell,m)$, we separate the wave function
$\psi(\vecr)$ as a product of the radial wave function $u_\ell(r)$ and spherical
harmonics $\YY_\ell^m(\theta,\phi)$,
\begin{equation}
 \psi(\vecr)=R_\ell(r)\YY_\ell^m(\theta,\phi)
 = \frac{u_\ell(r)}{r}\YY_\ell^m(\theta,\phi) \,.
\end{equation}
This leads to the radial Schr\"odinger equation
\begin{equation}
 \left[\frac{\dd^2}{\dd r^2} - \frac{\ell(\ell+1)}{r^2} - U(r)
 + p^2\right]u_\ell(r) = 0 \,,
\label{eq:SG-rad}
\end{equation}
with $p=\sqrt{2\mu E}=\sqrt{-2\mu E_B}\equiv\ii\kappa$ and $U(r)\equiv2\mu
V(r)$.  The mass is denoted by $\mu$ since in later applications it will be the
reduced mass for a two-particle system.  The normalized radial wave function
satisfies
\begin{equation}
 \int_0^\infty \dd r\,|u_\ell(r)|^2 = 1 \,.
\label{eq:NC-rad}
\end{equation}

For vanishing potential, we have two linearly independent solutions of the free
wave equation.  These are the Riccati--Bessel functions $\hat{\jmath}_\ell(pr)$
and Riccati--Neumann functions $\hat{n}_\ell(pr)$, which can be expressed in
terms of ordinary Bessel functions.  It is often more convenient to work with
the Riccati--Hankel functions
\begin{equation}
\hat{h}_\ell^\pm(z) = \hat{n}_\ell(z) \pm \ii\hat{\jmath}_\ell(z).
\end{equation}
These have the asymptotic form $\ee^{\pm\ii z}$ as $|z|\to\infty$.  For future
reference, we give the explicit expressions for $\hat{h}_\ell^+(z)$ for
$\ell=0,1,2$:
\begin{subequations}
\begin{align}
\label{eq:h-hat-0}
 \hat{h}_0^+(z) &= \ee^{\ii z} \,, \\
\label{eq:h-hat-1}
 \hat{h}_1^+(z) &= \left(1+\frac{\ii}{z}\right)\ee^{\ii(z-\nicefrac\pi2)} \,, \\
\label{eq:h-hat-2}
 \hat{h}_2^+(z) &= \left(1+\frac{3\ii}{z}-\frac{3}{z^2}\right)
 \ee^{\ii(z-\pi)} \,.
\end{align}
\end{subequations}

\subsubsection*{Asymptotic form of bound state wave functions}

In order to discuss the asymptotic form of bound state wave functions we define
solutions $\chi^\pm_{\ell,p}(r)$ of \eqref{eq:SG-rad} that fulfill the
condition
\begin{equation}
 \chi^\pm_{\ell,p}(r) \xrightarrow{r\to\infty} \hat{h}_\ell^\pm(pr) \,.
\end{equation}
The bound-state solution normalized according to \eqref{eq:NC-rad} can then be
written as
\begin{equation}
 u_\ell(r) = \ii^\ell\gamma\,\chi^+_{\ell,p}(r) \,,
\end{equation}
where $p=\ii\kappa$ and
\begin{equation}
 \gamma=\left(\int\nolimits_0^\infty\dd r\,|\chi^+_{\ell,p}(r)|^2\right)^{-1/2}
\label{eq:gamma}
\end{equation}
is the asymptotic normalization coefficient.\footnote{For more details, see
Ref.~\cite{Taylor}, Chaps. 11 and 12.}  The factor $\ii^\ell$ adjusts the phase
such that $u_\ell(r)$ is a real function.  If the potential has a finite range
$R$, \ie, $U(r)=0$ for $r>R$, we have the exact identity
\begin{equation}
 u_\ell(r) = \ii^\ell\gamma\,\hat{h}_\ell^+(\ii\kappa r)
 \mathtext{for} r > R \,.
\label{eq:u-asympt}
\end{equation}
\medskip

The asymptotic normalization $\gamma$ is itself an interesting quantity.  In
low-energy astrophysical reactions it determines the capture rate of a single
proton or neutron at zero relative energy~\cite{Xu:1994zz}.  In the shallow
binding limit it is also directly related to low-energy scattering parameters. 
For $\ell=0$,
\begin{subequations}
\label{eq:gamma-r}
\begin{equation}
 r_0 + \frac2{\gamma^2} - \frac1\kappa = \OO(\kappa)
 \mathtext{as} \kappa\to0 \,,
\label{eq:gamma-r-0}
\end{equation}
whereas for angular momenta $\ell\geq1$ one has
\begin{equation}
 r_\ell + \frac{2\kappa^{2\ell}}{\gamma^2} = \OO(\kappa)
 \mathtext{as} \kappa\to0 \,.
\label{eq:gamma-r-ell}
\end{equation}
\end{subequations}
We note that $\gamma$ will in general depend upon the binding 
momentum $\kappa$.  We will derive the relations~\eqref{eq:gamma-r-0}
and~\eqref{eq:gamma-r-ell} in Appendix~\ref{sec:ShallowBinding}.  The S-wave
formula is well known and given in~\cite{GbWatson}, although without the
$\OO(\kappa)$ correction estimate.  For higher angular momentum the situation is
qualitatively different due to the divergence of $\hat{h}_\ell^\pm(z)$ at $z =
0$.

\subsection{Infinite volume}
\label{sec:InfVol}

We now consider a system of two spinless particles with reduced mass $\mu$
with zero total momentum.  We will work with relative coordinates
$\vecr=\vec{r}_1-\vec{r}_2$.  The interaction is given by a potential
$V(\vecr,\vecr')$, which we in general allow to be non-local in configuration
space.  It is assumed to be Hermitian, rotationally symmetric, and to have a
finite range $R$, \ie,
\begin{equation}
 V(\vecr,\vecr') = 0 \mathtext{if} |\vecr|>R \mathtext{or} |\vecr'|>R \,.
\label{eq:V-FR}
\end{equation}

We consider the case when the system has a bound state $\ket{\psi_B}$ with
angular quantum numbers $(\ell,m)$.  The Schr\"odinger equation,
\begin{equation}
 \hat{H}\ket{\psi_B} = -{E_B}\ket{\psi_B} \,,
\label{eq:SG-op}
\end{equation}
can be written as
\begin{equation}
 -{\frac1{2\mu}}\Laplace_r\,\psi_B(\vecr)
 + \int\ddr'\,V(\vecr,\vecr')\,\psi_B(\vecr') = -{E_B}\,\psi_B(\vecr)
\label{eq:SG-nonlocal}
\end{equation}
in configuration space.  We note that for a local potential,
\begin{equation}
 V(\vecr,\vecr')=V(\vecr)\,\vdelta(\vecr-\vecr') \,,
\end{equation}
Eq.~\eqref{eq:SG-nonlocal} reduces to the familiar form
\begin{equation}
 \left[-{\frac1{2\mu}}\Laplace_r\,+ V(\vecr)\right]\,\psi_B(\vecr) =
-{E_B}\,\psi_B(\vecr) \,.
\label{eq:SG-local} 
\end{equation}
Regardless of the locality of the interaction, the wave function
$\psi_B(\vecr)$ has the asymptotic form
\begin{equation}
 \psi_B(\vecr) = \ii^\ell\gamma\,\YY_\ell^m(\vecr/r)\,
 \frac{\hat{h}_\ell^+(i\kappa r)}{r} \mathspace (r > R) \,,
\label{eq:psi-asympt}
\end{equation}
where $\displaystyle\kappa=\sqrt{2\mu E_B}$ is the binding momentum.

\subsection{Finite volume}
\label{sec:FinVol}

We now consider what happens when the two-body system is put into a cubic
periodic box with length $L \gg R$.  For this problem it is convenient to
define a periodic extension of the potential
\begin{equation}
 V_L(\vecr,\vecr') = \sum\limits_{\vecn\in\ZZ^3}
 V(\vecr+\vecn L,\vecr'+\vecn L) \,.
\end{equation}
We take $\ket{\psi}$ to be an exact periodic solution of the 
finite volume Schr\"odinger equation,
\begin{equation}
 \hat{H}_L\ket{\psi} = -E_B(L)\ket{\psi}\,,
\label{eq:SG-FV}
\end{equation}
with the finite-volume Hamiltonian $\hat{H}_L=\hat{H}_0+\hat{V}_L$.  It is clear
that $E_B(L)$ approaches the infinite volume eigenvalue $E_B$ and
$\ket{\psi}\to\ket{\psi_B}$ as $L\to\infty$.\medskip

We now derive a formula for the finite volume mass (energy) shift,
\begin{equation}
 \Delta m_B \equiv E_B(\infty) - E_B(L) \,.
\label{eq:Delta_m-direct}
\end{equation}
Let us define a state $\ket{\psi_0}$ by adding together periodic copies of the
infinite volume wave function in \eqref{eq:SG-nonlocal},
\begin{equation}
 \braket{\vecr}{\psi_0} = \psi_0(\vecr)
 = \sum\limits_{\vecn}\psi_B(\vecr+\vecn L) \,.
\end{equation}
This clearly satisfies the periodicity condition.  Acting upon this state with
the finite volume Hamiltonian, we get
\begin{equation}
\begin{split}
 H_L \psi_0(\vecr) &= H_0 \sum\limits_{\vecn'} \psi_B(\vecr+\vecn'L)
 + \sum\limits_{\vecn'}\sum\limits_{\vecn}\int\ddr'\,
 V(\vecr+\vecn L,\vecr'+\vecn L)\psi_B(\vecr'+\vecn'L) \\
 &= \sum\limits_{\vecn'}\Bigg\{H_0\,\psi_B(\vecr+\vecn'L)
 + \int\ddr'\,V(\vecr+\vecn'L,\vecr'+\vecn'L)\,\psi_B(\vecr'+\vecn'L) \\
 &\hspace{10em}+\sum\limits_{\vecn\neq\vecn'}\int\ddr'\,
 V(\vecr+\vecn L,\vecr'+\vecn L)\,\psi_B(\vecr'+\vecn'L)\Bigg\} \\
 &= -{E_B(\infty)} \sum\limits_{\vecn'} \psi_B(\vecr+\vecn'L)
 + \sum\limits_{\vecn'}\sum\limits_{\vecn\neq\vecn'}
 \int\ddr'\,V(\vecr+\vecn L,\vecr'+\vecn L)\,\psi_B(\vecr+\vecn'L) \,.
\end{split}
\end{equation}
The final result can be written as
\begin{equation}
 \hat{H}_L \ket{\psi_0} = -{E_B(\infty)} \ket{\psi_0} + \ket{\eta} \,,
\label{eq:H_L-psi_0}
\end{equation}
where we have defined $\ket{\eta}$ as
\begin{equation}
 \eta(\vecr) = \sum\limits_{\vecn'}\sum\limits_{\vecn\neq\vecn'}
 \int\ddr'\,V(\vecr+\vecn L,\vecr'+\vecn L)\,\psi_B(\vecr'+\vecn'L) \,.
\end{equation}
With the substitution $\vecr'\to\vecr'-\vecn L$ for each term in the sum, this
can be rewritten as
\begin{equation}
 \eta(\vecr) = \sum\limits_{\vecn'}\sum\limits_{\vecn\neq\vecn'}\int\ddr'\,
 V(\vecr+\vecn L,\vecr')\,\psi_B\big(\vecr'+(\vecn'-\vecn)L\big) \,.
\end{equation}
Due to the finite range of the potential we only get contributions from the
domain $|\vecr'|<R$.  We note that $|\vecr'+(\vecn'-\vecn)L|>R$ when
$\vecn\neq\vecn'$ and $R \ll L$.  Therefore we can use the asymptotic form of
the wave function and find that $\ket{\eta}=\OO\big(\ee^{-{\kappa L}}\big)$. 
This means that $\ket{\psi_0}$ is an approximate solution of the finite volume
Schr\"odinger equation~\eqref{eq:SG-FV} for large $L$.  Motivated by this, we
write the exact finite volume solution $\ket{\psi}$ explicitly as
\begin{equation}
 \ket{\psi} = \alpha\ket{\psi_0} + \ket{\psi'} \mathtext{with} \ket{\psi'}
 = \OO\big(\ee^{-{\kappa L}}\big) \,.
\label{eq:psi-prime}
\end{equation}
We take $\ket{\psi}$ to be unit-normalized per volume $L^3$.  The same is true
of $\ket{\psi_0}$ up to corrections of order $\ee^{-\kappa L}$.  We will choose
$\alpha$ such that
\begin{equation}
 \braket{\psi'}{\psi_0} = 0 \,.
\end{equation}
\medskip

Consider now the matrix element $\bra{\psi}{\hat{H}_L}\ket{\psi_0}$.  Acting
with $\hat{H}_L$ on $\ket{\psi_0}$, we get
\begin{equation}
 \bra{\psi}{\hat{H}_L}\ket{\psi_0}
 = -{E_B(\infty)}\braket{\psi}{\psi_0} + \braket{\psi}{\eta}
 = -{E_B(\infty)}\braket{\psi_0}{\psi_0}\cdot\alpha + \braket{\psi}{\eta}
\end{equation}
according to \eqref{eq:H_L-psi_0} and \eqref{eq:psi-prime}.  On the other hand,
acting with $\hat{H}_L$ on $\bra{\psi}$ yields
\begin{equation}
 \bra{\psi}{\hat{H}_L}\ket{\psi_0}
 = -{E_B(L)}\braket{\psi}{\psi_0}
 = -{E_B(L)}\braket{\psi_0}{\psi_0}\cdot\alpha \,.
\end{equation}
Combining these two results we find
\begin{equation}
 E_B(\infty)-E_B(L) = \Delta m_B
 = \frac{\braket{\psi}{\eta}}{\alpha\braket{\psi_0}{\psi_0}} \,.
\label{eq:Delta_m-raw}
\end{equation}

We first consider the numerator in this expression.  Obviously,
\begin{equation}
 \braket{\psi}{\eta} = \alpha\braket{\psi_0}{\eta} + \braket{\psi'}{\eta}
 = \alpha\braket{\psi_0}{\eta} + \OO\big(\ee^{-{2\kappa L}}\big) \,.
\end{equation}
We note that the factor of $\alpha$ here will cancel the $\alpha$ in the
denominator of~\eqref{eq:Delta_m-raw}.  We can now simplify further starting
with
\begin{equation}
 \braket{\psi_0}{\eta} = \sum\limits_{\vecn''}\sum\limits_{\vecn'}
 \sum\limits_{\vecn\neq\vecn'}\int\ddr\int\ddr'\,\psi_B^*(\vecr+\vecn''L)\,
 V(\vecr+\vecn L,\vecr'+\vecn L)\,\psi_B(\vecr'+\vecn'L) \,.
\end{equation}
For each $\vecn$ we can make the substitutions $\vecr\to\vecr-\vecn L$ and
$\vecr'\to\vecr'-\vecn L$.  These leave the integrals invariant, and we get
\begin{equation}
 \braket{\psi_0}{\eta} = \sum\limits_{\vecn''}\sum\limits_{\vecn'}
 \sum\limits_{\vecn\neq\vecn'}\int\ddr\int\ddr'\,
 \psi_B^*\big(\vecr+(\vecn''-\vecn)L\big)\,V(\vecr,\vecr')\,
 \psi_B\big(\vecr'+(\vecn'-\vecn)L\big) \,.
\end{equation}
Setting $\vecm=\vecn'-\vecn$ and $\vecm'=\vecn''-\vecn$ yields
\begin{equation}
 \braket{\psi_0}{\eta} = C\cdot\sum\limits_{\vecm'}\sum\limits_{\vecm\neq\vZero}
 \int\ddr\int\ddr'\,\psi_B^*(\vecr+\vecm'L)\,V(\vecr,\vecr')\,
 \psi_B(\vecr'+\vecm L) \,,
\label{eq:psi0-eta}
\end{equation}
where $C$ counts the number of repeated periodic copies.  The fact that $C$
diverges simply reflects the fact that we are working with periodic wave
functions with normalization measured per volume $L^3$, and $C$ will cancel in
the final result.  For the integral to be non-zero, both $\vecr$ and $\vecr'$
have to be close to $\vZero$ due to the finite range of the potential.  From the
assumption $L \gg R$ it then follows that all terms with $\vecm'\neq\vZero$ are
suppressed by at least a factor of $\ee^{-{2\kappa L}}$, and we have
\begin{equation}
\braket{\psi_0}{\eta} = C\cdot\sum\limits_{\vecm\neq\vZero}
 \int\ddr\int\ddr'\,\psi_B^*(\vecr)\,V(\vecr,\vecr')\,\psi_B(\vecr'+\vecm L)
 + \OO\big(\ee^{-{2\kappa L}}\big) \,.
\end{equation}
The possible nonvanishing values of $|\vecm|$ are $1,\sqrt2,\sqrt3,\ldots\;$.
We therefore arrive at
\begin{equation}
 \braket{\psi_0}{\eta} = C\cdot\sum\limits_{|\vecm|=1}\int\ddr\int\ddr'\,
 \psi_B^*(\vecr)\,V(\vecr,\vecr')\,\psi_B(\vecr'+\vecm L)
 + \OO\big(\ee^{-{\sqrt2\kappa L}}\big) \,.
\label{eq:psi0-eta-final}
\end{equation}
For the denominator in \eqref{eq:Delta_m-raw}, an analogous procedure yields
\begin{equation}
 \braket{\psi_0}{\psi_0} = C\cdot\sum\limits_{\vecm}\int\ddr\,
 \psi_B^*(\vecr)\,\psi_B(\vecr+\vecm L)
 = C\cdot\left[1+\OO\big(\ee^{-{\kappa L}}\big)\right]
\label{eq:psi0-psi0-final}
\end{equation}
with the same constant $C$ as above.  Combining~\eqref{eq:psi0-eta-final}
and~\eqref{eq:psi0-psi0-final}, the constant $C$ cancels and we get
\begin{equation}
 \Delta m_B = \sum\limits_{|\vecn|=1}\int\ddr\int\ddr'\,
 \psi_B^*(\vecr)\,V(\vecr,\vecr')\,\psi_B(\vecr'+\vecn L)
  + \OO\big(\ee^{-{\sqrt2\kappa L}}\big) \,,
\label{eq:Delta_m-int}
\end{equation}
where we have renamed $\vecm$ back to $\vecn$.\medskip

Eq.~\eqref{eq:Delta_m-int} is a general result valid for any angular momentum.
The dependence of the mass shift upon quantum numbers $(\ell,m)$ will emerge
from the wave function $\psi_B$ and the resulting overlap integrals
in~\eqref{eq:Delta_m-int}.  In the following, we explore this dependence in
detail and denote the mass shift as $\Delta m_B^{(\ell,m)}$.

\section{L\"uscher's result for S-waves}
\label{sec:S-wave}

For $\ell=0$ the asymptotic wave function~\eqref{eq:psi-asympt} is given as
\begin{subequations}
\begin{equation}
 \psi_B(\vecr) = \psi_B(|\vecr|) = \sqrt{\frac1{4\pi}}\frac{u_0(r)}{r}
\end{equation}
with
\begin{equation}
 u_0(r) = \gamma\,\hat{h}_0^+(\ii\kappa r) = \gamma\,\ee^{-{\kappa r}}
 \mathtext{for} r > R \,.
\end{equation}
\end{subequations}
Due to the finite range $R \ll L$ of the potential we only have contributions
with $|\vecr'+\vecn L| > R$ in~\eqref{eq:Delta_m-int}.  Hence we can insert the
asymptotic form for $\psi_B(\vecr'+\vecn L)$ and get
\begin{equation}
 \Delta m_B^{(0,0)} = \frac\gamma{\sqrt{4\pi}}\sum\limits_{|\vecn|=1}
 \int\ddr\int\ddr'\,\psi_B^*\big(|\vecr|\big)\,V(\vecr,\vecr')\,
 \frac{\ee^{-\kappa|\vecr'+\vecn L|}}{|\vecr'+\vecn L|}
 + \OO\big(\ee^{-{\sqrt2\kappa L}}\big) \,.
\end{equation}
We can furthermore use the Schr\"odinger equation~\eqref{eq:SG-nonlocal} to
eliminate the potential.  Doing this and then renaming $\vecr'\to\vecr$, we get
\begin{equation}
\begin{split}
 \Delta m_B^{(0,0)} &= \frac\gamma{\sqrt{4\pi}}\sum\limits_{|\vecn|=1}
 \int\ddr\left\{\left[\frac{\Laplace_{r}}{2\mu}-E_B\right]
 \psi_B^*\big(|\vecr|\big)\right\}
 \frac{\ee^{-\kappa|\vecr+\vecn L|}}{|\vecr+\vecn L|}
 + \OO\big(\ee^{-{\sqrt2\kappa L}}\big) \\
 &= \frac\gamma{\sqrt{4\pi}}\sum\limits_{|\vecn|=1}
 \int\ddr\,\psi_B^*\big(|\vecr-\vecn L|\big)
 \frac{1}{2\mu}\left[\Laplace_r-\kappa^2\right]\frac{\ee^{-\kappa r}}{r}
 + \OO\big(\ee^{-{\sqrt2\kappa L}}\big) \,.
\end{split}
\end{equation}
In the second line we have shifted the integration variable and used partial
integration to let the Laplacian act on $\exp(-\kappa r)/r$.  Finally, we use
the fact that $\exp(-\kappa r)/(4\pi r)$ is a Green's function for the operator
$\Laplace_r-\kappa^2$,
\begin{equation}
 \left[\Laplace_r-\kappa^2\right]\frac{\ee^{-\kappa r}}{4\pi r}
 = -\vdelta(\vecr) \,.
\label{eq:Helmholtz-GF}
\end{equation}
This allows us to perform the integral and arrive at
\begin{equation}
\begin{split}
 \Delta m_B^{(0,0)} &= -\frac{\sqrt\pi\gamma}\mu\sum\limits_{|\vecn|=1}
 \psi_B^*\big(|\vecn L|\big) + \OO\big(\ee^{-{\sqrt2\kappa L}}\big) \\
 &= -{3}|\gamma|^2\,\frac{\ee^{-{\kappa L}}}{\mu L}
 + \OO\big(\ee^{-{\sqrt2\kappa L}}\big).
\label{eq:Delta_m-S}
\end{split}
\end{equation}
In the last step we have inserted the asymptotic form of the wave function for
$\psi_B^*\big(|\vecn L|\big) = \psi_B^*(L)$, and the sum yields a factor of six.
This is just L\"uscher's result~\eqref{eq:Delta_m-Luscher} as given in the
introduction.

\section{Extension to higher partial waves}
\label{sec:P-wave-and-higher}

We now discuss the generalization of the mass shift formula to arbitrary
angular momentum.  The general form for the asymptotic wave function is
\begin{equation}
 \psi_{B,(\ell,m)}(\vecr) = Y_\ell^m(\theta,\phi)\,
 \frac{\ii^\ell\gamma\hat{h}_\ell^+(\ii\kappa r)}{r} \,.
\label{eq:psi-asympt-general}
\end{equation}
Inserting this into~\eqref{eq:Delta_m-int} and performing steps analogous to
those presented above for the S-wave case, we find
\begin{equation}
 \Delta m_B = \sum\limits_{|\vecn|=1}\int\ddr\,\left\{\frac1{2\mu}\!
 \left[\Laplace_r-\kappa^2\right]\!\psi_B^*(\vecr-\vecn L)\right\}\,
 \YY_\ell^m(\theta,\phi)\frac{\ii^\ell\gamma\hat{h}_\ell^+(\ii\kappa r)}r
 + \OO\big(\ee^{-{\sqrt2\kappa L}}\big)\,.
\label{eq:Delta_m-int-asympt}
\end{equation}

The crucial ingredient is the relation
\begin{equation}
 Y_\ell^m(\theta,\phi)\,\frac{\hat{h}_\ell^+(\ii\kappa r)}{r}
 = \left(-\ii\right)^\ell R_\ell^m\left(-\frac1\kappa\vNabla_r\right)
 \left[\frac{\ee^{-\kappa r}}{r}\right] \,,
\label{eq:SH-Hankel}
\end{equation}
where $R_\ell^m$ are the solid harmonics defined via
$R_\ell^m(x,y,z) = R_\ell^m(\vecr) = r^\ell Y_\ell^m(\theta,\phi)$.  The 
derivation of \eqref{eq:SH-Hankel} follows from Lemma B.1 in
Ref.~\cite{Luscher:1990ux}, which proves that
\begin{equation}
 R_\ell^m(\vNabla)f(r) = R_\ell^m(\vecr)\left(\frac1r\ddd{r}\right)^\ell f(r)
\label{eq:Lemma}
\end{equation}
for any smooth function $f(r)$.  We obtain~\eqref{eq:SH-Hankel} by using the
relation\footnote{The relation~\eqref{eq:HankelDeriv} is just a special case of
Eq.~(10.1.24) in~\cite{AbramStegPocket}, which also holds for other spherical
Bessel functions.},
\begin{equation}
 \left(\frac1z\frac{\dd}{\dd z}\right)^{\!\ell} h^{(1)}_0(z)
 = (-1)^\ell\,z^{-\ell}\,h^{(1)}_\ell(z)\,,
\label{eq:HankelDeriv}
\end{equation}
and noting that $\ee^{-\kappa r}=\hat{h}_0^+(\ii\kappa r)$ and
$\hat{h}^+_\ell(z) = \ii z\,h^{(1)}_\ell(z)$,  where $h^{(1)}_\ell(z)$ is a
spherical Hankel function of the first kind.

We can illustrate \eqref{eq:SH-Hankel} using an example.  For the case 
$\ell=1$ and $m=0$ we have
\begin{equation}
 \hat{h}_1^+(\ii\kappa r)
 \sim \left(1+\frac1{\kappa r}\right)\frac{\ee^{-\kappa r}}{r}
\end{equation}
and $\YY_1^0(\theta,\phi) \propto \cos\theta$.  A straightforward calculation
shows that indeed
\begin{equation}
 \cos\theta\left(1+\frac1{\kappa r}\right)\frac{\ee^{-\kappa r}}{r}
 = -\frac1\kappa\del{z}\left[\frac{\ee^{-\kappa r}}{r}\right] \,,
\end{equation}
with $\cos\theta=z/r$.

Using~\eqref{eq:SH-Hankel} to rewrite~\eqref{eq:Delta_m-int-asympt}, we get
\begin{equation}
 \Delta m_B = \frac{\gamma}{2\mu}\sum\limits_{|\vecn|=1}\int\ddr
 \left\{\left[\Laplace_r-\kappa^2\right]\!\psi_B^*(\vecr-\vecn L)
 \vphantom{\left[\frac{\ee^{-\kappa r}}{r}\right]}\right\}
 \left\{R_\ell^m\left(-\frac1\kappa\vNabla_r\right)
 \left[\frac{\ee^{-\kappa r}}{r}\right]\right\}
 + \OO\big(\ee^{-{\sqrt2\kappa L}}\big) \,.
\end{equation}
We now integrate by parts and pass the Laplacian through the differential
operator $R_\ell^m(-\vNabla_r/\kappa)$.  Since the operators both consist of
partial derivatives, this is not a problem when the wave function is smooth.  We
assume that this is the case, with the possible exception of a measure zero
region that can be omitted from the integral.\medskip

The partial integrations give a factor $(-1)^\ell$.  We can now proceed in
exactly the same way as for S-waves.  We perform one more integration by parts
so that the Laplacian acts on $\exp(-\kappa r)/r$.  This yields a delta function
times a factor of $-4\pi$, and the final result is then
\begin{equation}
 \Delta m_B^{(\ell,m)} =
 (-1)^{\ell+1}\cdot\frac{2\pi\gamma}\mu\sum\limits_{|\vecn|=1}
 R_\ell^m\left(-\frac1\kappa\vNabla_r\right)\psi^*_{B,(\ell,m)}(\vecr-\vecn L)
 \Bigg|_{\vecr=\vZero} + \OO\big(\ee^{-{\sqrt2\kappa L}}\big) \,.
\label{eq:Delta_m-general}
\end{equation}
For $\psi^*_{B,(\ell,m)}$ we can insert the asymptotic
form~\eqref{eq:psi-asympt-general} since it is evaluated in the asymptotic
region.

\subsection{Results}
\label{sec:Results}

For $\ell=1$, we find the same result for all three P-wave states,
\begin{equation}
 \Delta m_B^{(1,0)} = \Delta m_B^{(1,\pm1)}
 = 3|\gamma|^2\,\frac{\ee^{-{\kappa L}}}{\mu L}
 + \OO\big(\ee^{-{\sqrt2\kappa L}}\big) \,.
\label{eq:Delta_m-P}
\end{equation}
When compared to the S-wave case, the sign of the P-wave mass shift is reversed
while the magnitude is the same.  Qualitatively, this means that S-wave bound
states are more deeply bound when put in a finite volume while P-wave bound
states are less bound.  This will be discussed in more detail later.\medskip

We next discuss the results for $\ell=2$. From~\eqref{eq:Delta_m-general}, we
find
\begin{align}
 \Delta m_B^{(2,0)} &= -15|\gamma|^2\,\frac{\ee^{-{\kappa L}}}{\mu L}
 \cdot F_2^0\!\left(\tfrac1{\kappa L}\right)
  + \OO\big(\ee^{-{\sqrt2\kappa L}}\big) \,, \\
 \Delta m_B^{(2,\pm1)} &= +15|\gamma|^2\,\frac{\ee^{-{\kappa L}}}{\mu L}
 \cdot F_2^1\!\left(\tfrac1{\kappa L}\right)
 + \OO\big(\ee^{-{\sqrt2\kappa L}}\big) \,, \\
 \Delta m_B^{(2,\pm2)} &= -15|\gamma|^2\,\frac{\ee^{-{\kappa L}}}{\mu L}
 \cdot F_2^2\!\left(\tfrac1{\kappa L}\right)
 + \OO\big(\ee^{-{\sqrt2\kappa L}}\big) \,,
\end{align}
where
\begin{align}
 F_2^0(x) &= \frac12 + 3x + \frac{27}2x^2 + \frac{63}2x^3+ \frac{63}2x^4 \,,\\
 F_2^1(x) &= 2x + 9x^2 + 21x^3  + 21x^4 \,, \\
 F_2^2(x) &= \frac14 + \frac12x + \frac94x^2 + \frac{21}4x^3 + \frac{21}4x^4 \,.
\end{align}

We note that the size and even the sign of the mass shift depends on the quantum
number $m$.  To understand this effect, we need to take into account that our
cubic finite volume breaks the rotational symmetry group down to a cubic
subgroup.

\subsubsection*{Representations of the cubic group}
\label{sec:Cubic}

The cubic symmetry group $\OO$ is a finite subgroup of $SO(3)$ with 24 elements.
There are five irreducible representations of $\OO$.  They are conventionally
called $A_1$, $A_2$, $E$, $T_1$ and $T_2$, and their dimensionalities are $1$,
$1$, $2$, $3$ and $3$, respectively.  Irreducible representations $D^\ell$ of
the rotation group $SO(3)$ are reducible with respect to $\OO$ for $\ell > 1$. 
For further details about the decomposition see, for example,
Ref.~\cite{Bernard:2008ax}.  

In our discussion we assume that the infinite volume system has no partial wave
mixing, such that orbital angular momentum $\ell$ is a good quantum number.  We
also assume that there are no accidental degeneracies in the bound state
spectrum, so we can use $\ell$ as a label for the family of cubic
representations split apart at finite volume.  Parity invariance remains
unbroken by the cubic volume, and we have $P=(-1)^\ell$ just as in the infinite
volume case.  For clarity, however, we will indicate parity explicitly with
$\pm$ superscripts in the following.

With our assumptions, an S-wave state in infinite volume will map onto an
$A_1^+$ state at finite volume.  Also a P-wave triplet will map onto the three
elements of the $T_1^-$ representation at finite volume.  For D-waves, however,
the five D-wave states are split into a $T_2^+$ triplet and an $E^+$ doublet,
\begin{equation}
 D^2 = T_2^+ \oplus E^+ \,.
\end{equation}

In the following we use the notation $\ket{\Gamma,\ell;\alpha}$,
$\alpha=1,\ldots,\dim(\Gamma)$, for the basis vectors of the irreducible cubic
representations.  We can rewrite the finite volume  mass shift in
Eq.~\eqref{eq:Delta_m-int} as
\begin{equation}
 \Delta m_B^{(\Gamma,\ell,\alpha)}
 \equiv \mbraketbig{\Gamma,\ell;\alpha}{\hat V\sum\limits_{|\vecn|=1}
 \hat T(\vecn L)}{\Gamma,\ell;\alpha} \,,
\end{equation}
where $\hat T(\vecx)$ is the translation operator by displacement $\vecx$.  We
can also calculate the matrix elements of $\Delta m_B$ in the $(\ell,m)$ basis.
In this case there will be off-diagonal matrix elements connecting $(\ell,m)$
and $(\ell,m^{\prime})$ when $m$ and $m^{\prime}$ are equivalent modulo 4.

According to Ref.~\cite{Bernard:2008ax}, the unitary transformation between 
the two basis sets for the five D-wave states is
\begin{subequations}
\begin{align}
 \ket{T_2^+,2;1} &= -{\tfrac1{\sqrt2}}\left(\ket{2,-1}+\ket{2,1}\right) \,,\\
 \ket{T_2^+,2;2}
 &= \phantom-{\tfrac\ii{\sqrt2}}\left(\ket{2,-1}-\ket{2,1}\right) \,, \\
 \ket{T_2^+,2;3} &= -{\tfrac1{\sqrt2}}\left(\ket{2,-2}-\ket{2,2}\right)
\end{align}
\end{subequations}
and
\begin{subequations}
\begin{align}
 \ket{E^+,2;1} &= \ket{2,0} \,, \\
 \ket{E^+,2;2} &= {\tfrac1{\sqrt2}}\left(\ket{2,-2}+\ket{2,2}\right) \,.
\end{align}
\end{subequations}
So, for example, we have
\begin{equation}
\begin{split}
 \Delta m_B^{(T_2^+,2;1)}
 &= \frac12\left(\Delta m_B^{(2,-1,-1)} + 2\Delta m_B^{(2,-1,1)}
 +\Delta m_B^{(2,1,1)}\right) \\
 &= -15|\gamma|^2\,\frac{\ee^{-{\kappa L}}}{\mu L}\cdot\left(\tfrac2{\kappa L}
 +\tfrac9{\kappa^2L^2}+\tfrac{21}{\kappa^3L^3}+\tfrac{21}{\kappa^4L^4}\right)
 + \OO\big(\ee^{-{\sqrt2\kappa L}}\big) \,,
\end{split}
\end{equation}
where we have defined
\begin{equation}
 \Delta m_B^{(\ell,m_1,m_2)} =
 (-1)^{\ell+1}\cdot\frac{2\pi\gamma}\mu\sum\limits_{|\vecn|=1}
 R_\ell^{m_1}\left(-\frac1\kappa\vNabla_r\right)
 \psi^*_{B,(\ell,m_2)}(\vecr-\vecn L)\Bigg|_{\vecr=0} \\
 + \OO\big(\ee^{-{\sqrt2\kappa L}}\big)
\label{eq:Delta_m-most-general}
\end{equation}
as a straightforward generalization of~\eqref{eq:Delta_m-general}.\medskip

As expected from cubic symmetry, the mass shift is the same for all three
$T_2^+$ states, and the same for both $E^+$ states.  To summarize our results,
we write the mass shift for a state belonging to irreducible representation
$\Gamma$ with angular momentum $\ell$ as
\begin{equation}
 \Delta m_B^{(\ell,\Gamma)} = \alpha\left({\tfrac{1}{\kappa L}}\right)
 \cdot|\gamma|^2\,\frac{\ee^{-{\kappa L}}}{\mu L}.
 + \OO\big(\ee^{-{\sqrt2\kappa L}}\big)
\label{eq:Delta_m-Gamma}
\end{equation}
We list the coefficients $\alpha\left(\tfrac{1}{\kappa L}\right)$ for
$\ell=0,\ldots,3$ in Table~\ref{tab:Results}.

\begin{table}[htbp]
 \begin{tabular}{c|c||c}
  $\ell$ & $\Gamma$ & $\alpha(x)$ \\
  \hline\hline
  $0$ & $A_1^+$ & $-3$ \\
  \hline
  $1$ & $T_1^-$ & $+3$ \\
  \hline
  $2$ & $T_2^+$ &
  $30x+135x^2+315x^3+315x^4$ \\
  $2$ & $E^+$ &
  $-\nicefrac{1}{2}\left(15+90x+405x^2+945x^3+945x^4\right)$ \\
  \hline
  $3$ & $A_2^-$ & $315x^2+2835x^3+12285x^4+28350x^5+28350x^6$ \\
  $3$ & $T_2^-$ & $-\nicefrac{1}{2}\left(105x+945x^2+5355x^3
   +19530x^4+42525x^5+42525x^6\right)$ \\
  $3$ & $T_1^-$ & $\quad\nicefrac{1}{2}\left(14+105x+735x^2+3465x^3
   +11340x^4+23625x^5+23625x^6\right)$ \\
 \end{tabular}
\caption{Coefficient $\alpha(x)$ in the expression for the finite volume mass
shifts for $\ell=0,\ldots,3$. $\Gamma$ indicates the corresponding
representation of the cubic group.}
\label{tab:Results}
\end{table}

\subsection{Sign of the mass shift}
\label{sec:Sign}

\begin{figure}[htbp]
\centering
\includegraphics[width=0.5\textwidth,clip]{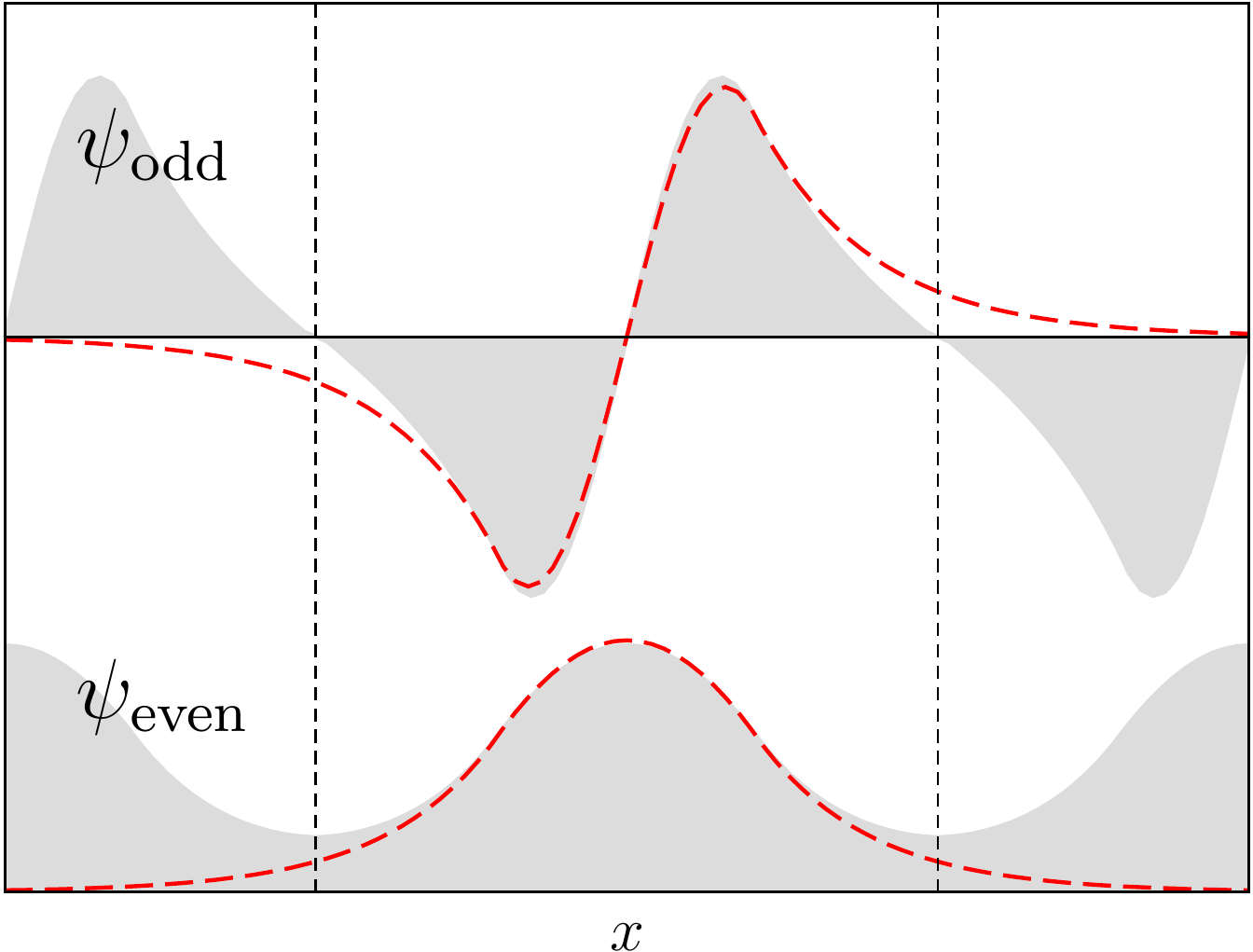}
\caption{Wave functions with even (bottom) and odd parity (top) for a
one-dimensional square well potential in a box with periodic boundary
conditions.  The dashed lines give the infinite volume solutions for
comparison.}
\label{fig:WF}
\end{figure}

The sign of the finite volume mass shift can be understood in terms of the
parity of the wave function.  In infinite volume the tail of each bound state
wave function must vanish at infinity.  In the finite volume, however, the bound
state wave functions with even parity along a given axis can remain nonzero
everywhere.  Only the derivative needs to vanish, and the kinetic energy is
lowered by broadening the wave function profile.  On the other hand, a wave
function with odd parity along a given axis must change sign across the
boundary.  In this case the wave function profile is compressed and the kinetic
energy is increased.  We have illustrated both cases for a one-dimensional
square well potential in Fig.~\ref{fig:WF}.

\medskip
In three dimensions, the situation is slightly more complicated, which can be
seen from the fact that for $\ell=2$ the sign of the mass shift depend on the
representation of the cubic group even though the parity is just $(-1)^2=+1$ for
all states.  In order to understand this, we consider the basis polynomials
for the cubic representations.  These basis polynomials are obtained by
decomposing the cubic basis vectors in terms of solid harmonics which are
homogeneous polynomials in $x$, $y$ and $z$.  For $\ell=0,\ldots,4$ the basis
polynomials are also given explicitly in~\cite{Luscher:1990ux}.\medskip

For a given polynomial $P(x,y,z)$, we define its \textit{leading parity} as
\begin{equation}
 \mathrm{lp}\,P = (-1)^{d_\mathrm{max}} \,,
\end{equation}
where
\begin{equation}
 d_\mathrm{max} = \max\{\deg_x P,\deg_y P,\deg_z P\}
\end{equation}
is the maximum degree of $P$ with respect to any one of the three variables.  It
is this leading parity that determines the asymptotic behavior of the mass shift
as $\kappa L \to \infty$.  More precisely, we have
\begin{equation}
 \alpha\left(\tfrac{1}{\kappa L}\right)
 \sim (-1)^{d_\mathrm{max}+1}
 \left(\tfrac{1}{\kappa L}\right)^{\ell-d_\mathrm{max}}
 \mathtext{as} \kappa L \to \infty
\end{equation}
for the $\alpha\left(\tfrac{1}{\kappa L}\right)$ in Eq.~\ref{eq:Delta_m-Gamma}.

It can easily be checked that this relation holds for all results presented in
Table~\ref{tab:Results}.  For $\ell=2$, for example, we have the basis
polynomials
\begin{subequations}
\begin{align}
 P_{2,T_2^+} &\sim xy\ ,\ yz\ ,\ zx \,, \\
 P_{2,E^+}   &\sim x^2-y^2\ ,\ y^2-z^2 \,,
\end{align}
\end{subequations}
and hence $d_\mathrm{max}=1$ for the $T_2^+$ representation and
$d_\mathrm{max}=2$ for the $E^+$ representation.

\subsection{Trace formula}
\label{sec:Trace}

The expressions for the finite volume mass shift become simpler when we sum
over all $m$ for a given $\ell$.  We can rewrite \eqref{eq:Delta_m-general} as
\begin{equation}
 \Delta m_B^{(\ell,m)} =
 (-1)^{\ell+1}\cdot\frac{2\pi\gamma}\mu\sum\limits_{|\vecn|=1}
 R_\ell^m\left(-\frac1\kappa\vNabla_r\right)\psi^*_{B,(\ell,m)}(\vecr)
 \Bigg|_{\vecr=\vecn L} \\
 + \OO\big(\ee^{-{\sqrt2\kappa L}}\big) \,.
\end{equation}
Inserting the asymptotic form of the wave function,
\begin{equation}
 \psi^*_{B,(\ell,m)}(\vecr)\Big|_{\vecr=\vecn L}
 = \left[\YY_\ell^m(\theta,\phi)
 \frac{\ii^\ell\gamma\,\hat{h}_\ell^+(\ii\kappa r)}r\right]^*
 \Bigg|_{\vecr=\vecn L} \,,
\end{equation}
and using \eqref{eq:SH-Hankel} a second time yields
\begin{equation}
 \Delta m_B^{(\ell,m)} =
 (-1)^{\ell+1}\cdot\frac{2\pi|\gamma|^2}\mu\sum\limits_{|\vecn|=1}
 R_\ell^m\left(-\frac1\kappa\vNabla_r\right)R_\ell^{*m}
 \left(-\frac1\kappa\vNabla_r\right)\left[\frac{\ee^{-\kappa r}}{r}\right]
 \Bigg|_{\vecr=\vecn L} \\
 + \OO\big(\ee^{-{\sqrt2\kappa L}}\big) \,.
\end{equation}
Now, from the well-known relation
\begin{equation}
 \sum_{m=-\ell}^\ell \YY_\ell^m(\theta,\phi)\YY_\ell^{*m}(\theta,\phi)
 = \frac{2\ell+1}{4\pi}
\end{equation}
and $R_\ell^m(\vecr) = r^\ell Y_\ell^m(\theta,\phi)$ we get an analogous
expression for the solid harmonics, which then carries over to
\begin{equation}
 \sum_{m=-\ell}^\ell R_\ell^m\left(-\frac1\kappa\vNabla_r\right)R_\ell^{*m}
 \left(-\frac1\kappa\vNabla_r\right)f(r)
 = \frac1{\kappa^{2\ell}}\cdot\frac{2\ell+1}{4\pi}\,(\Laplace_r)^\ell f(r)
\end{equation}
for any sufficiently smooth function $f(r)$.  Finally, we have
\begin{equation}
 (\Laplace_r)^\ell\,\frac{\ee^{-\kappa r}}{r}
 = \kappa^{2\ell}\,\frac{\ee^{-\kappa r}}{r} \mathspace (r\neq0) \,,
\label{eq:Laplace-GF}
\end{equation}
which follows from Eq.~\eqref{eq:Helmholtz-GF}. Putting everything
together, we arrive at
\begin{equation}
\begin{split}
 \sum_{m=-\ell}^\ell \Delta m_B^{(\ell,m)}
 &= (-1)^{\ell+1}\cdot\frac{2\pi|\gamma|^2}\mu\cdot\frac{2\ell+1}{4\pi}
 \sum\limits_{|\vecn|=1}\frac1{\kappa^{2\ell}}\,
 (\Laplace_r)^\ell\left[\frac{\ee^{-\kappa r}}{r}\right]
 \Bigg|_{\vecr=\vecn L} + \OO\big(\ee^{-{\sqrt2\kappa L}}\big) \\
 &= (-1)^{\ell+1}(2\ell+1)\cdot3|\gamma|^2\frac{\ee^{-\kappa L}}{\mu L}
 + \OO\big(\ee^{-{\sqrt2\kappa L}}\big)\,,
\end{split}
\label{eq:avshift-1}
\end{equation}
where the sum just yields a factor of six.  Dividing by $2\ell +1$, we obtain
the average mass shift for states with angular momentum $\ell$,
\begin{equation}
 \Delta m_B^{(\ell)}=(-1)^{\ell+1}\cdot3|\gamma|^2\frac{\ee^{-\kappa L}}{\mu L}
 + \OO\big(\ee^{-{\sqrt2\kappa L}}\big) \,.
\label{eq:avshift-2}
\end{equation}
Apart from the alternating sign, this average shift is independent of $\ell$.
\medskip

Eq.~\eqref{eq:avshift-2} can be verified explicitly for the the results
presented in Sec.~\ref{sec:Results} (\cf~Table~\ref{tab:Results}).  For $\ell=2$,
for example, one has to average over the three-dimensional representation
$T_2^+$ and the two-dimensional representation $E^+$.\footnote{The mapping from
the angular momentum eigenstates to the cubic group states is a unitary
transformation.}

\section{Numerical tests}
\label{sec:NumTest}

In order to verify our predictions numerically, we put the Schr\"odinger
equation~\eqref{eq:SG-FV} on a discrete spatial lattice such that the
Hamiltonian becomes an ordinary matrix.  We then calculate the corresponding
energy eigenvalues and eigenvectors.

\subsection{Lattice discretisation}
\label{sec:Lattice}

We use a hat symbol to denote dimensionless lattice units.  For example, we have
\begin{equation}
 \hat L = L / a \mathtext{and} \hat E_B = E_B \cdot a \,,
\end{equation}
where $a$ denotes the lattice spacing.\medskip

The free lattice Hamiltonian is given by
\begin{equation}
 \hat H_0 = \sum\limits_{\hat\vecn}\left[
 \frac3{\hat\mu}\,a^\dagger(\hat\vecn)a(\hat\vecn)
 - \frac1{2\hat\mu}\sum\limits_{l=1,2,3}
 \left(a^\dagger(\hat\vecn)a(\hat\vecn+\hat\vece_l)
 +a^\dagger(\hat\vecn)a(\hat\vecn-\hat\vece_l)\right)\right]
\label{eq:H-latt}
\end{equation}
where $a^\dagger(\hat\vecn)$ and $a(\hat\vecn)$ are creation and annihilation
operators for a lattice site $\hat\vecn$ and $\hat\vece_l$ is a unit vector in
the $l$-direction.  The corresponding lattice dispersion relation is
\begin{equation}
 \hat E(\hat\vecq)=\frac{Q^2(\hat\vecq)}{2\hat\mu}
\label{eq:LattDisp}
\end{equation}
with the lattice function
\begin{equation}
 Q^2(\hat\vecq) = 2\sum\limits_{l=1,2,3}(1-\cos\hat q_i)
 = \sum\limits_{l=1,2,3}\hat q_l^2\left[1+\OO(\hat q_l^2)\right]
\end{equation}
and the lattice momenta
\begin{equation}
 \hat\vecq = 2\pi{\hat\vecn}/{\hat L} \,.
\end{equation}
The binding momentum for a bound state with energy $-\hat E_B$ is determined by
\begin{equation}
 {-\hat\mu\hat E_B} = \left(1-\cos(-\ii\hat\kappa)\right)
 = \left(1-\cosh(\hat\kappa)\right) \,.
\label{eq:LattDisp-BS}
\end{equation}
The lattice Green's function for the Hamiltonian~\eqref{eq:H-latt} is
\begin{equation}
 \hat G(\hat\vecn,\hat E) = \frac1{L^3}\sum\limits_{\hat\vecq}
 \frac{\ee^{-\ii\hat\vecq\cdot\hat\vecn}}{Q^2(\hat\vecq)+2\hat\mu\hat E} \,.
\end{equation}

We impose periodic boundary conditions by defining the distance $\hat r$ to the
origin as
\begin{equation}
 \hat r\left(\hat\vecn\right) = \sqrt{\sum\limits_{l=1,2,3}
 \min\left\{\hat n_l^2, \left( \hat L - \hat n_l \right)^2\right\}} \,.
\end{equation}

\subsection{Methods}
\label{sec:Methods}

We calculate the mass shift using three different methods:
\begin{enumerate}
 \item As a direct difference in energies, Eq.~\eqref{eq:Delta_m-direct}, where
we use a very large volume ($L_\infty$) to approximate the infinite volume
result.
 \item From the overlap formula \eqref{eq:Delta_m-int}.
 \item Using discretized versions of \eqref{eq:Delta_m-S}
and~\eqref{eq:Delta_m-P}, which we obtain by replacing $\exp(-\kappa r)/r$ with
the lattice Green's function.  More precisely, we write the asymptotic bound
state wave function~\eqref{eq:psi-asympt} as
\begin{equation}
 \psi_B(\vecr) = \ii^\ell\gamma\,\YY_\ell^m(\vecr/r)\,\hat{h}_\ell^+(i\kappa r)
 \,\ee^{\kappa r}\cdot4\pi G_\kappa(r) \mathspace (r > R)
\end{equation}
and replace the continuum Green's function
\begin{equation}
 G_\kappa(\vecr) = \frac{\ee^{-\kappa r}}{4\pi r}
\end{equation}
with the lattice version
\begin{equation}
 \hat G_{\hat\kappa}(\hat\vecn)
 \equiv \hat G\left(\hat\vecn,\frac{-\hat\kappa^2}{2\hat\mu}\right) \,.
\end{equation}
Effectively, this amounts to the replacement
\begin{equation}
 {\ee^{-\hat\kappa \hat L}}/{\hat L}
 \longrightarrow 4\pi\hat G_{\hat\kappa}(\hat L,0,0)
\end{equation}
in the mass shift formula.
\end{enumerate}

The lattice Green's function is also used to calculate the asymptotic
normalization $\gamma$ from the lattice data.  This procedure has the advantage
of avoiding large lattice discretization errors.

\subsection{Results}
\label{sec:ResultsNum}

In the following we report physical quantities in units where the reduced mass
$\mu$ is set to 1.

\subsubsection{Gaussian potential}
\label{sec:GaussPot}

\begin{figure}[htbp]
\centering
\includegraphics[width=0.5\textwidth,clip]{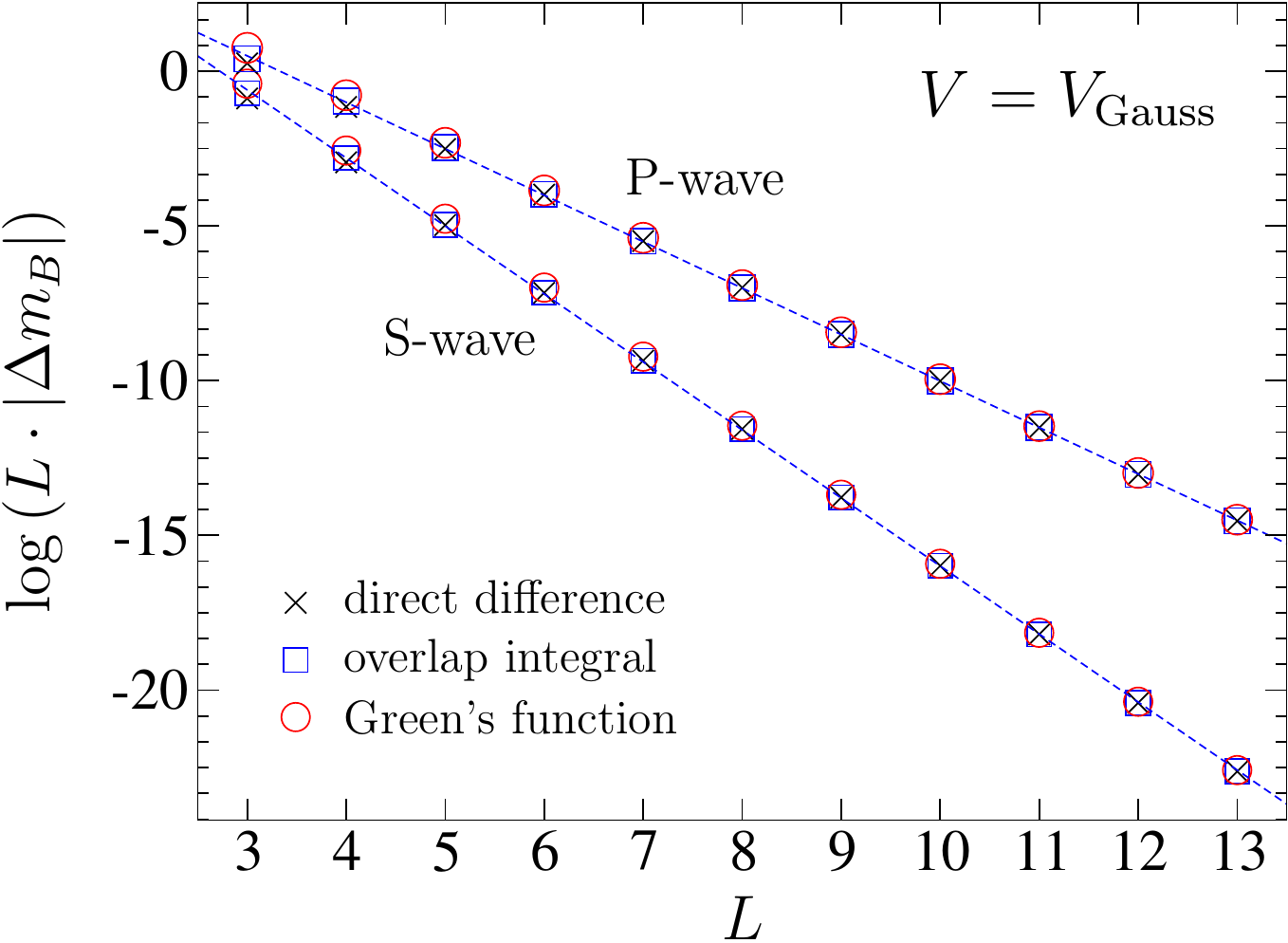}
\caption{S-wave and P-wave mass shifts $\log(L\cdot|\Delta m_B|)$ as functions
of the box size $L$ (in lattice units) for a Gaussian potential.
We show the results obtained from the direct difference
Eq.~\eqref{eq:Delta_m-direct} (crosses), evaluation of the overlap integral
Eq.~\eqref{eq:Delta_m-int} (squares), and discretized versions of
Eqs.~\eqref{eq:Delta_m-S}, \eqref{eq:Delta_m-P} (circles).  The dashed lines
show linear fits to the overlap integral results.}
\label{fig:Shifts-SP-Gauss}
\end{figure}

We first use a Gaussian potential,
\begin{equation}
 V_\text{Gauss}(r) = -V_0\,\exp\left(-r^2/(2R^2)\right) \,,
\end{equation}
with $R=1$ and $V_0 = 6$.  This potential does not have a finite range in a
strict mathematical sense, but the range corrections can be entirely neglected
in comparison with other errors in our numerical calculation.  The smoothness of
the Gaussian potential helps to minimize lattice discretization artifacts.  In
Fig.~\ref{fig:Shifts-SP-Gauss} we show the S- and P-wave mass shifts obtained
with the three methods described in Sec.~\ref{sec:Methods} above.  The results
from the three different methods described above agree well for both S- and
P-waves.  In order to compare the dependence on the box size $L$ with the
predicted behavior, we have plotted $\log(L\cdot|\Delta m_B|)$ against $L$ (we
use the absolute value of $\Delta m_B$ since the S-wave mass shift is negative).
For both S- and P-waves, the expected linear dependence is clearly visible.

When we perform a linear fit to the overlap integral data (dashed lines in
Fig.~\ref{fig:Shifts-SP-Gauss}) we obtain $\kappa = 2.198 \pm 0.005$, $|\gamma|
= 11.5 \pm 0.2$ for the S-wave results and $\kappa = 1.501 \pm 0.004$, $|\gamma|
= 7.0 \pm 0.1$ for the P-wave results.  The values for the asymptotic
normalization are in good agreement with the results $|\gamma|\sim 11.5$
(S-wave) and $|\gamma| \sim 7.2$ (P-wave) that are obtained directly from the
$L_\infty=40$ data.  Inserting the corresponding energy eigenvalues into the
lattice dispersion relation~\eqref{eq:LattDisp-BS}, we find $\kappa \sim 2.211$
(S-wave) and $\kappa \sim 1.501$ (P-wave), again in quite good agreement with
the fit results.  The remaining small discrepancies can be attributed to the
mixing with higher partial waves induced by the lattice discretization and the
fact that we have not performed a continuum extrapolation to vanishing lattice
spacing.

\subsubsection{Simple step potential}
\label{sec:StepPot}

\begin{figure}[htbp]
\centering
\includegraphics[width=0.5\textwidth,clip]{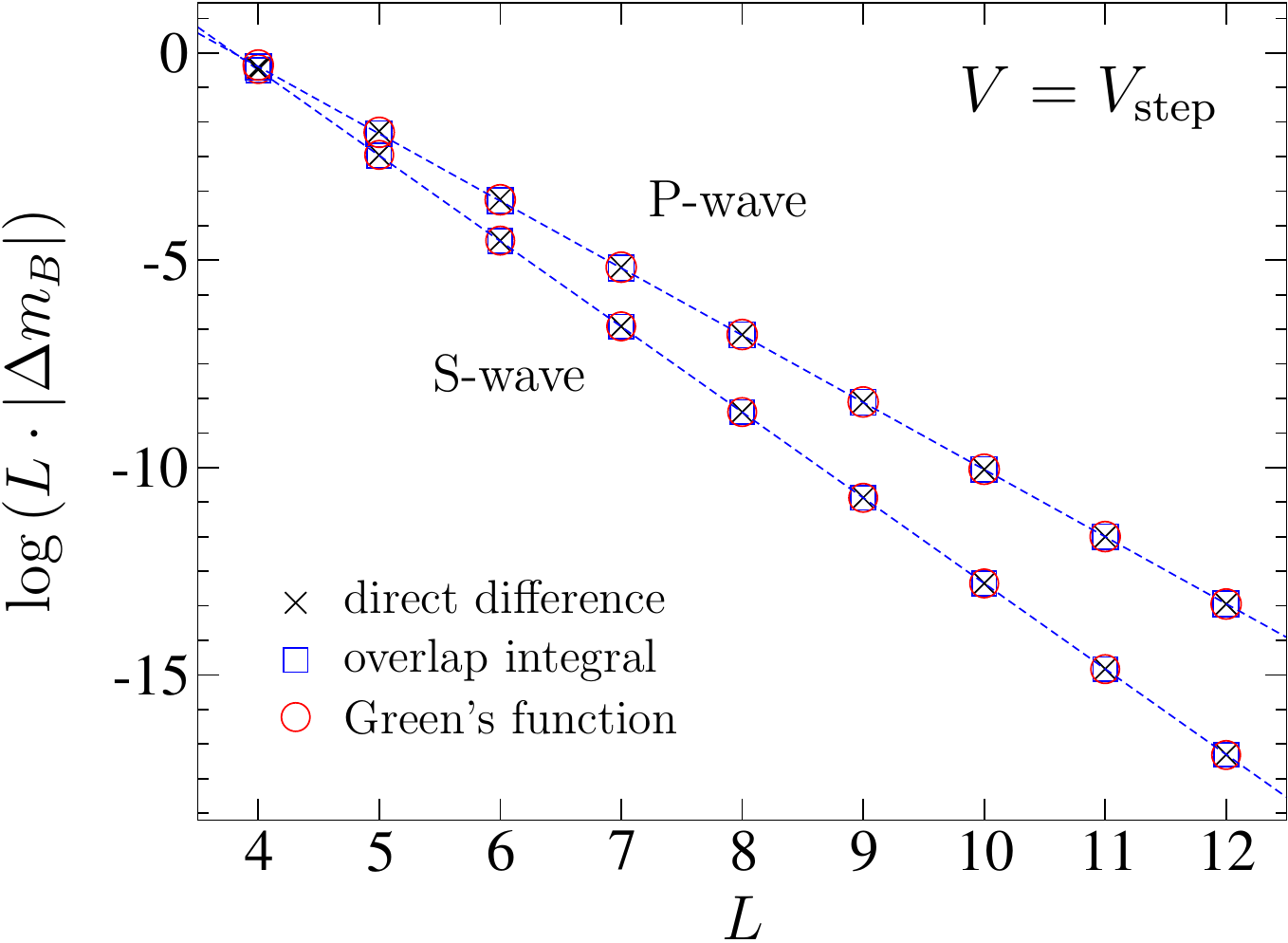}
\caption{S-wave and P-wave mass shifts $\log(L\cdot|\Delta m_B|)$ as functions
of the box size $L$ (in lattice units) for a simple step potential.  The symbols
are as in Fig.~\ref{fig:Shifts-SP-Gauss}.}
\label{fig:Shifts-SP-Step}
\end{figure}

For a simple step potential,
\begin{equation}
 V_\text{step}(r) = -V_0\,\theta(R-r) \,,
\end{equation}
which we use with $R=2$ and $V_0=3$, the numerical calculation becomes more
difficult since the discontinuous shape introduces considerable lattice
artifacts.  Yet we discuss it here because it has a strict finite range and we
find that for a small lattice spacing of $a=0.2$ the results are satisfactory.
In Fig.~\ref{fig:Shifts-SP-Step}, we show a plot analogous to the one presented
for the Gaussian potential.  Again, the results from the different methods
agree well and the expected linear behavior is clearly visible.  Furthermore,
the results from the three methods agree well with each other already for
smaller $L$ (compared to the results for the Gaussian potential), as expected
from the fact that the step potential does not have a tail.

From fitting to the overlap integral data (dashed lines in
Fig.~\ref{fig:Shifts-SP-Step}) we obtain $\kappa = 2.0636 \pm 0.0005$, $|\gamma|
= 29.17 \pm 0.06$ for the S-wave results and $\kappa = 1.6192 \pm 0.0009$,
$|\gamma| = 12.48 \pm 0.05$ for the P-wave results.  From the $L_\infty=40$ data
we find $\kappa \sim 2.0666$, $|\gamma|\sim 29.6$ (S-wave) and $\kappa \sim
1.6242$, $|\gamma|\sim 12.8$ (P-wave).  Given that we do not have error
estimates for the $L_\infty=40$ results, the overall agreement is quite good.
\medskip

\begin{figure}[htbp]
\centerline{
\includegraphics[width=0.49\textwidth,clip]{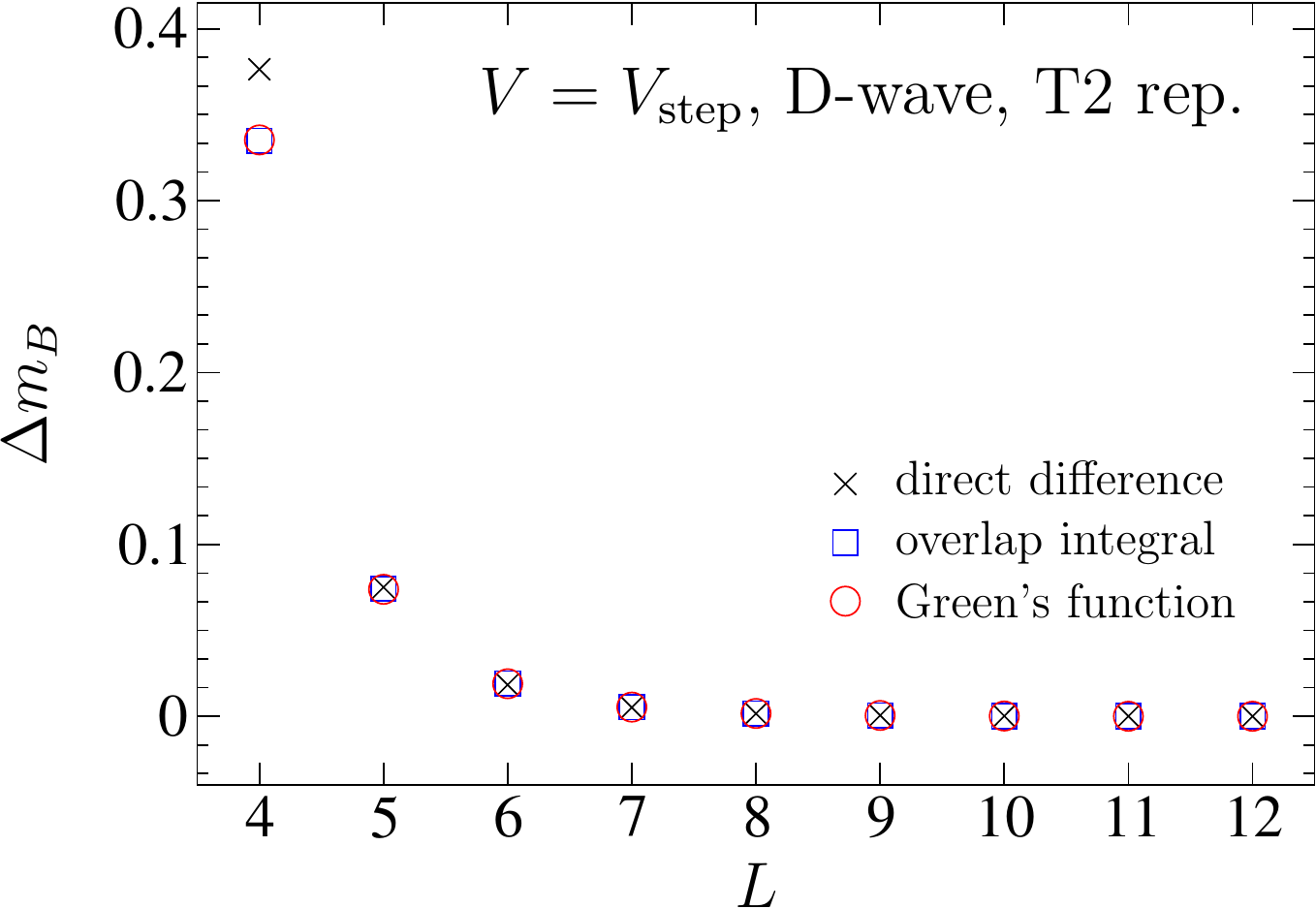}
\;
\includegraphics[width=0.49\textwidth,clip]{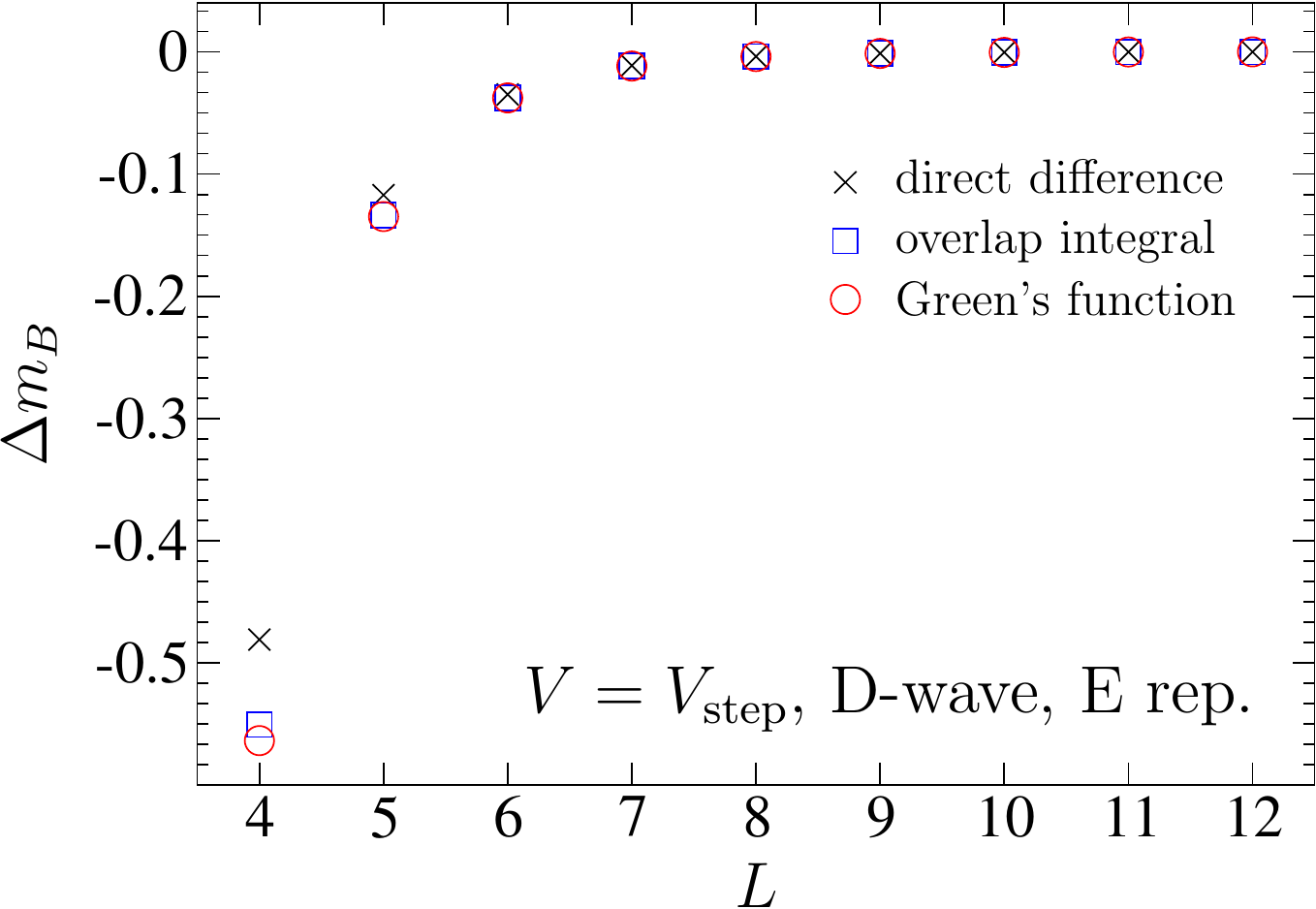}}
\caption{D-wave, mass shift  $\Delta m_B$ for $T_2^+$ rep. (left panel) and
$E^+$ rep. (right panel) as a function of the box size $L$ (in lattice units)
for a simple step potential.
The symbols are as in Fig.~\ref{fig:Shifts-SP-Gauss}.}
\label{fig:Shift-D-T2-E-Step}
\end{figure}

Finally, we also check our result for the D-wave mass splittings using the step
potential with $a=0.2$.  In Fig.~\ref{fig:Shift-D-T2-E-Step}, we show the mass
shift for the D-wave states in both the $T_2^+$ and the $E^+$ representation.
Due to the polynomial coefficients $\alpha\left(\tfrac{1}{\kappa L}\right)$ (see
Eq.~\eqref{eq:Delta_m-Gamma} and Table~\ref{tab:Results}) one does not expect a
linear dependence on $L$ for $\log(L\cdot|\Delta m_B|)$.  Hence, we simply plot 
$\Delta m_B$ as a function of $L$ directly and do not perform a fit.
Nevertheless, we see that (except for very small $L$, where obviously the
condition $L\gg R$ is not satisfied) the agreement between the three methods to
calculate $\Delta m_B$ is very good and hence conclude that our mass shift
formula indeed gives the right result also for $\ell=2$.

\section{Two-dimensional systems}
\label{sec:2D}

In this section we derive a formula for the finite-volume (or rather
finite-area) mass shift of bound states in two-dimensional systems. The results
can be used, for example, in lattice simulations of cold atomic systems, which
can be prepared experimentally to be effectively
two-dimensional~\cite{Dyke:2011,Frohlich:2011}. \medskip

In two dimensions, the Schr\"odinger equation is
\begin{equation}
 -{\frac1{2\mu}}\Laplace_r^{\mathrm{2D}}\,\psi_B(\vecr)
 + \int\dd^2r'\,V(\vecr,\vecr')\,\psi_B(\vecr') = -{E_B}\,\psi_B(\vecr)
\label{eq:SG-nonlocal-2D}
\end{equation}
with
\begin{equation}
 \Laplace_r^{\mathrm{2D}}\,\psi_B(\vecr)
 = \left[\frac1r\del{r}+\frac{\partial^2}{\partial r^2}
 + \frac1{r ^2}\frac{\partial^2}{\partial\theta^2}\right]\psi_B(\vecr)
\end{equation}
in polar coordinates.  States are described by a single angular momentum quantum
number $m=0,\pm1,\pm2,\ldots$, and for the wave function we have the separation
\begin{equation}
 \psi_B(\vecr) = u_{m}(r) Y_m(\theta)
\end{equation}
with
\begin{equation}
  Y_m(\theta) = \frac{\ee^{\ii m \theta}}{\sqrt{2\pi}}\ \,.
\label{eq:SH-2D}
\end{equation}
The two linearly independent solutions of the free radial equation
\begin{equation}
 \left(\frac{\dd^2}{\dd r^2} +\frac1r\frac{\dd}{\dd r}
 - \frac{m^2}{r^2} + p^2\right) u_m(r) = 0 \,,
\label{eq:SG-rad-free-2D}
\end{equation}
are just the Bessel and Neumann functions $J_m(pr)$ and $N_m(pr)$.  For a bound
state, we have $p^2 = -\kappa^2 = -2\mu E_B$, and the wave function has the
asymptotic form
\begin{equation}
 u_m(r) = \gamma\,K_m(\kappa r) \mathtext{for} r > R \,,
\label{eq:u-asympt-2D}
\end{equation}
where $K_m$ is the modified Bessel function of the second kind.  It is related
to the Hankel function of the first kind,
\begin{equation}
 H_m^{(1)}(z) = J_m(z) + \ii N_m(z) \,,
\end{equation}
via
\begin{equation}
 K_m(x) = \frac\pi2 \ii^{m+1} H_m^{(1)}(\ii x) \,.
\label{eq:ModBess}
\end{equation}
As in the three-dimensional case, $\gamma$ is the asymptotic normalization.
Inserting~\eqref{eq:ModBess} into~\eqref{eq:u-asympt-2D} yields a form which is
more similar to the three-dimensional expression, Eq.~\eqref{eq:u-asympt}.  To
make the analogy to the calculations in Sec.~\ref{sec:P-wave-and-higher} as
explicit as possible, we will use the Hankel function in the following
intermediate steps and only express the final results in terms of the modified
Bessel function.\medskip

Nearly all of the three-dimensional calculation carries over and we just replace
all exponential terms by Hankel functions.  The overlap integral for the mass
shift is now
\begin{equation}
 \Delta m_B^{(m)} = \sum\limits_{|\vecn|=1}\int\dd^2r\int\dd^2r'\,
 \psi_{B,m}^*(\vecr)\,V(\vecr,\vecr')\,\psi_{B,m}(\vecr'+\vecn L)
  + \OO\left(\ii H_m^{(1)}\big(\sqrt2\kappa L\big)\right) \,.
\label{eq:Delta_m-int-2D}
\end{equation}
From the asymptotic form of the Hankel function,
\begin{equation}
 H_m^{(1)}(z)
 \sim \sqrt{\frac2{\pi z}}\,\ee^{\ii\left(z-\frac{m}2\pi-\frac\pi4\right)}
 \mathtext{as} |z| \to \infty \,,
\end{equation}
it is clear that in principle we still have an exponential behavior.  In
deriving Eq.~\eqref{eq:Delta_m-int-2D} we have used this to write
\begin{equation}
 \OO\left(H_m^{(1)}\big(\ii\kappa L\big)^2\right)
 \sim \OO\left(H_m^{(1)}\big(2\ii\kappa L\big)\right)
 \ll \OO\left(H_m^{(1)}\big(\sqrt2\ii\kappa L\big)\right) \,.
\end{equation}
In the following, we will simply write the correction terms as
$\OO\big(\ee^{-{\sqrt2\kappa L}}\big)$, as in the three-dimensional
case.\medskip

The two-dimensional analogue of the relation~\eqref{eq:SH-Hankel} is
\begin{equation}
 Y_m(\theta)H_m^{(1)}(\ii\kappa r)
 = (-\ii)^m R_m\left(-\frac1{\kappa}\vNabla^{\mathrm{2D}}\right)
 H_0^{(1)}(\ii\kappa r) \,,
\label{eq:SH-Hankel-2D}
\end{equation}
where $R_m(r,\theta)=r^m Y_m(\theta)$.  This follows from
\begin{equation}
 R_m(\vNabla^{\mathrm{2D}})f(r) =
 R_m(\vecr)\left(\frac1r\ddd{r}\right)^m f(r)
\label{eq:Lemma-2D}
\end{equation}
and
\begin{equation}
 \left(\frac1z\frac{\dd}{\dd z}\right)^{\!m} H^{(1)}_0(z)
 = (-1)^m\,z^{-m}\,H^{(1)}_m(z)
\label{eq:HankelDeriv-2D}
\end{equation}
The proof for Eq.~\eqref{eq:Lemma-2D} can be carried out in the same manner as
the three-dimensional proof in Lemma B.1 in~\cite{Luscher:1990ux}, where one
uses the expansion of $\ee^{\ii\vecp\cdot\vecr}$ (2D vectors) in terms of Bessel
functions.  As the final ingredient, we have
\begin{equation}
 \left[\Laplace_r^{\mathrm{2D}}-\kappa^2\right]
 \frac{\ii}4 H_0^{(1)}(\ii\kappa r) = -\delta^{(2)}(\vecr) \,.
\label{eq:Helmholtz-GF-2D}
\end{equation}
Using all this in steps completely analogous to those in three dimensions, we
get
\begin{equation}
 \Delta m_B^{(m)} =
 (-1)^{m+1}\cdot\frac{\pi\gamma}\mu\sum\limits_{|\vecn|=1}
 R_m\left(-\frac1\kappa\vNabla_r^{\mathrm{2D}}\right)\psi^*_{B,m}
 (\vecr-\vecn L) \Bigg|_{\vecr=0} \\
 + \OO\big(\ee^{-{\sqrt2\kappa L}}\big) \,.
\label{eq:Delta_m-general-2D}
\end{equation}
For $m=0$ (two-dimensional S-waves), this directly yields
\begin{equation}
 \Delta m_B^{(0)} = -2 \frac{|\gamma|^2}{\mu} K_0(\kappa L)
 + \OO\big(\ee^{-{\sqrt2\kappa L}}\big) \,.
\end{equation}

In fact, Eq.~\eqref{eq:Delta_m-general-2D} can be simplified further.  Inserting
the asymptotic form for the wave function for $\psi^*_{B,m}$ and
using~\eqref{eq:SH-Hankel-2D} a second time gives
\begin{multline}
 \Delta m_B^{(m)} =
 (-1)^{m+1}\cdot\frac{\pi|\gamma|^2}\mu\sum\limits_{|\vecn|=1}
 R_m\left(-\frac1\kappa\vNabla_r^{\mathrm{2D}}\right)
 R_m^*\left(-\frac1\kappa\vNabla_r^{\mathrm{2D}}\right)
 \left[\ii\frac\pi2 H_0^{(1)}(\ii\kappa r)\right]
 \Bigg|_{\vecr=\vecn L} \\
 + \OO\big(\ee^{-{\sqrt2\kappa L}}\big) \,.
\end{multline}
From~\eqref{eq:SH-2D} and $R_m(\vecr) = r^mY_m(\theta)$ it is clear that
\begin{equation}
 R_m(\theta) R_m^*(\theta) = \frac{(r^2)^m}{2\pi} \,,
\end{equation}\
which then yields
\begin{equation}
 R_m\left(-\frac1\kappa\vNabla_r^{\mathrm{2D}}\right)
 R_m^*\left(-\frac1\kappa\vNabla_r^{\mathrm{2D}}\right) f(r)
 = \frac1{\kappa^{2m}}\cdot\frac{1}{2\pi}\,
 \left(\Laplace_r^{\mathrm{2D}}\right)^m f(r)
\end{equation}
for any sufficiently smooth $f(r)$.  This is essentially the same relation that
we used to derive the trace formula in the three-dimensional case, only that
here we do not have to sum over different $m$.  Together with the
two-dimensional analogue of~\eqref{eq:Laplace-GF},
\begin{equation}
 \left(\Laplace_r^{\mathrm{2D}}\right)^m H_0^{(1)}(\ii\kappa r)
 = \kappa^{2m}H_0^{(1)}(\ii\kappa r) \mathspace (r\neq0) \,,
\label{eq:Laplace-GF-2D}
\end{equation}
we then get
\begin{equation}
\begin{split}
 \Delta m_B^{(m)}
 &= (-1)^{m+1}\cdot\frac{\gamma|^2}{2\mu}
 \sum\limits_{|\vecn|=1}\frac1{\kappa^{2m}}\,
 \left(\Laplace_r^{\mathrm{2D}}\right)^m
 \left[\ii\frac\pi2 H_0^{(1)}(\ii\kappa r)\right]\Bigg|_{\vecr=\vecn L}
 + \OO\big(\ee^{-{\sqrt2\kappa L}}\big) \\
 &= (-1)^{m+1}\cdot\frac{2|\gamma|^2}{\mu} K_0(\kappa L)
 + \OO\big(\ee^{-{\sqrt2\kappa L}}\big) \,.
\end{split}
\label{eq:Delta_m-general-2D-2}
\end{equation}
As we shall see in the following, this is the final result for $m=0$ and any odd
$m$, whereas for even $m\neq0$ things become slightly more complicated.\medskip

In general, we have to take into account that the finite volume breaks the
original planar rotational symmetry of the system down to the symmetry group of
a square.  We find that states with the same absolute value of $m$ may mix to
form good eigenstates in the finite volume.  More precisely, we have the
symmetric and antisymmetric combinations
\begin{equation}
 \ket{m,\pm} = \frac1{\sqrt2}\big(\ket{m}\pm\ket{-m}\big) 
\end{equation}
for $m\neq0$.  When we calculate the mass shift for these states (in the same
way as described in Sec.~\ref{sec:Results}), we get mixing terms of the form
\begin{multline}
 \Delta m_B^{(m,\mathrm{mixed})} =
 (-1)^{m+1}\cdot\frac{\pi\gamma}\mu\sum\limits_{|\vecn|=1}
 R_m\left(-\frac1\kappa\vNabla_r^{\mathrm{2D}}\right)\psi^*_{B,{-m}}
 (\vecr) \Bigg|_{\vecr=\vecn L}
 + \OO\big(\ee^{-{\sqrt2\kappa L}}\big) \,.
\label{eq:Delta_m-mixed-2D}
\end{multline}

Since the condition for the mixing of states is
\begin{equation}
 2m \equiv 0 \mod 4 \,,
\end{equation}
they do not play a role for odd $m$ (in fact, they vanish in this case).  For
even $m$, however, we have to take them into account and find
\begin{equation}
 \Delta m_B^{(m,\pm)} = \frac12\left(
 \Delta m_B^{(m)} \pm 2\Delta m_B^{(m,\mathrm{mixed})}+\Delta m_B^{(-m)}\right)
\end{equation}
as our final result.  As an illustration, we give the explicit results for
$|m|=2$:
\begin{subequations}
\begin{equation} 
 \Delta m_B^{(2,+)} = -4\frac{|\gamma|^2}\mu\left[
 \left(1+\tfrac{12}{\kappa^2L^2}\right)K_0(\kappa L)
 +\left(\tfrac4{\kappa L}+\tfrac{24}{\kappa^3L^3}\right)K_1(\kappa L)\right]
 + \OO\big(\ee^{-{\sqrt2\kappa L}}\big) \,,
\end{equation}
\begin{equation} 
 \Delta m_B^{(2,-)} = 16\frac{|\gamma|^2}\mu\left[
 \tfrac3{\kappa^2L^2}K_0(\kappa L)
 +\left(\tfrac1{\kappa L}+\tfrac6{\kappa^3 L^3}\right)\right]
 + \OO\big(\ee^{-{\sqrt2\kappa L}}\big) \,.
\end{equation}
\end{subequations}

\section{Summary and outlook}
\label{sec:Conclusion}

In this work we have derived explicit formulae for the mass shift of P- and
higher-wave bound states in a finite volume and discussed their decomposition
into states transforming according to the representations of the cubic group.
We have compared our numerical results for $\ell \leq 2$ with numerical
calculations of the finite-volume dependence for lattice Gaussian and step
potentials and found good agreement with the predictions.  For $\ell \geq 2$,
the mass shift of a given state $(\ell,m)$ depends on the angular momentum
projection $m$ due to the breaking of rotational symmetry.  When we average
over all $m$ in a multiplet, however, the absolute value of the mass shift is
even independent of $\ell$.  The mass shift for states in representations of the
cubic group, however, is the same for all states. The sign of the mass shift can
be understood from the leading parity of the representations.  Finally, we have
derived corresponding expressions for the finite volume mass shift in
two-dimensional systems.

Lattice calculations also provide a method to extract asymptotic normalization
coefficients, which are of interest in low-energy astrophysical capture 
reactions.  We have pointed out how the asymptotic normalization and binding 
momentum of a shallow bound state can be used to extract the effective range
from a simulation.

Our work provides a general framework for future lattice studies of molecular
states with angular momentum in systems with short-range interactions.
Applications to nuclear halo systems and molecular states in atomic and hadronic
physics appear promising.  An important next step would be to include Coulomb
effects into the framework in order to investigate the volume dependence of
bound states of charged particles, which are much easier to treat
experimentally.  In particular, this extension of the formalism is important if
one wants to describe proton-halo nuclei.  Another interesting direction would
be to analyze the volume dependence of resonances along the lines of
Refs.~\cite{Luscher:1991cf,Bernard:2008ax}.

\begin{acknowledgments}
This research was supported in part by the DFG through SFB/TR 16 ``Subnuclear
structure of matter'', the BMBF under contract No. 06BN9006, and by the US
Department of Energy under contract No. DE-FG02-03ER41260.
S.K. was supported by the ``Studien\-stiftung des deutschen Volkes'' and by the
Bonn-Cologne Graduate School of Physics and Astronomy.
\end{acknowledgments}

\appendix

\section{Shallow binding limit}
\label{sec:ShallowBinding}

In this section, we derive the relation between the asymptotic normalization
of the bound state wave function and the effective range of the corresponding
two-particle scattering process that we have quoted in Sec.~\ref{sec:Prereq}.
\medskip

It was already pointed out by L\"uscher in~\cite{Luscher:1985dn} that the
asymptotic normalization $\gamma$ of the bound state wave function is related to
scattering parameters.  More precisely, the analytically-continued elastic
scattering amplitude in forward direction has a pole at the bound state energy, 
and the residue of this pole is proportional to $|\gamma|^2$.

In the limit of shallow bound states, \ie, $\kappa\to0$, it is possible to make
a more direct connection to the effective range in the corresponding scattering
channel.  The crucial ingredients for this are given in a general investigation
of Wigner causality bounds~\cite{Hammer:2010fw}.  We express the elastic
scattering amplitude $f_\ell(p)$ in terms of the scattering phase shift
$\delta_\ell(p)$,
\begin{equation}
 f_\ell(p) \propto
\frac{p^{2\ell}}{p^{2\ell+1}\left[\cot\delta_\ell(p)-\ii\right]} \,,
\end{equation}
for which we have the well-known effective range expansion
\begin{equation}
 p^{2\ell+1}\cot\delta_\ell(p)=-\frac1{a_\ell}+\frac12r_\ell\,p^2 + \cdots \,,
\label{eq:ERE}
\end{equation}
where $a_\ell$ and $r_\ell$ are the scattering and effective range parameters,
respectively.  A scattering state is described by a wave function
$u_\ell^{(p)}(r)$ that is a solution of the radial Schr\"odinger
equation~\eqref{eq:SG-rad} for positive real center-of-mass momentum $p>0$.  As
it is done in~\cite{Hammer:2010fw}, we choose the normalization such that
outside the range of the potential ($r>R$) we have
\begin{equation}
 u_\ell^{(p)}(r) = p^\ell
 \left[\cot\delta_\ell(p)\,\hat{\jmath}_\ell(pr)+\hat{n}_\ell(pr)\right] \,.
\end{equation}
For a bound state, we have $p=\ii\kappa$ and $\cot\delta_\ell(p)=\ii$.  Hence,
the wave function is
\begin{equation}
 u_\ell^{(\kappa)}(r) = (\ii\kappa)^\ell
 \left[\ii\hat{\jmath}_\ell(\ii\kappa r)+\hat{n}_\ell(\ii\kappa r)\right] \\
 = \ii^\ell \kappa^\ell \hat{h}_\ell^+(\ii\kappa r)
\end{equation}
in this case.  We immediately see that
\begin{equation}
 u_\ell^{(\kappa)}(r) = \frac{\kappa^\ell}{\gamma} u_\ell(r) \,,
\end{equation}
where $u_\ell(r)$ is the radial bound state wave function from the previous
sections (normalized to $1$).  Note that the phase convention is chosen such
that the wave functions are real.  From~\cite{Hammer:2010fw} we have the
relation
\begin{equation}
 r_\ell = b_\ell(R)
 - 2\lim_{\kappa\to0}\int_0^R\dd r\left[u_\ell^{(\kappa)}(r)\right]^2
\label{eq:ER-1}
\end{equation}
with
\begin{equation}
\begin{split}
 b_\ell(r) &=
 -\frac{2\Gamma\left(\ell-\frac12\right)\Gamma\left(\ell+\frac12\right)}\pi
 \left(\frac R2\right)^{\!-2\ell+1} \\
 &\hspace{10em}-\frac4{\ell+\frac12}\frac1{a_\ell}\left(\frac R2\right)^2
 +\frac{2\pi}{\Gamma\left(\ell+\frac32\right)\Gamma\left(\ell+\frac52\right)}
 \frac1{a_\ell^2}\left(\frac R2\right)^{\! 2\ell+3} \\
 &= -\frac{2\Gamma\left(\ell-\frac12\right)\Gamma\left(\ell+\frac12\right)}\pi
 \left(\frac R2\right)^{\!-2\ell+1} + \OO(a_\ell^{-1}) \,.
\label{eq:b-ell}
\end{split}
\end{equation}
For $\ell=0$, we get
\begin{equation}
 b_0 = 2R + \OO(a_\ell^{-1}) \,.
\label{eq:b-0}
\end{equation}
Note that $a_\ell^{-1}\to0$ as $\kappa\to0$, and since we only consider
finite-range potentials, we can in fact write
$\OO(a_\ell^{-1})=\OO(\kappa)$.\medskip

Using the normalization of the wave function we can rewrite~\eqref{eq:ER-1} as
\begin{equation}
 r_\ell = b_\ell(R)
 - 2\lim_{\kappa\to0}\left\{\frac{\kappa^{2\ell}}{\gamma^2}
 - \int_R^\infty\dd r\left[u_\ell^{(\kappa)}(r)\right]^2\right\} \,.
\label{eq:ER-2}
\end{equation}
The remaining integral can be expressed as
\begin{equation}
 \int_R^\infty\dd r\left[u_\ell^{(\kappa)}(r)\right]^2
 = \kappa^{2\ell}\int\limits_R^\infty\mathrm{d}r\,
 \left[\ii^\ell\hat{h}_\ell^+(\ii\kappa r) \right]^2 \,.
\label{eq:int-exterior-intermediate}
\end{equation}
For $\ell=0$, we get
\begin{equation}
 \int_R^\infty\dd r\left[u_\ell^{(\kappa)}(r)\right]^2
 = \frac{\ee^{-2\kappa R}}{2\kappa}
 = \frac1{2\kappa} - R + \OO(\kappa) \mathtext{as} \kappa\to0 \,.
\end{equation}
Together with~\eqref{eq:b-0} this yields
\begin{equation}
 r_0 + \frac2{\gamma^2} - \frac1\kappa = \OO(\kappa) \,.
\end{equation}
Up to the given order this is equivalent to the relation
\begin{equation}
 \gamma^2 = \frac{2\kappa}{1-\kappa r_0}
\end{equation}
from~\cite{GbWatson}.  For $\ell\geq1$, the integral is
\begin{equation}
 \int_R^\infty\dd r\left[u_\ell^{(\kappa)}(r)\right]^2
 = \frac{\Gamma\left(\ell-\frac12\right)\Gamma\left(\ell+\frac12\right)}\pi
 \left(\frac R2\right)^{-2\ell+1} + \OO(\kappa)
 \mathtext{as} \kappa\to0 \,.
\label{eq:int-exterior}
\end{equation}
We see that the leading term exactly cancels the one in~\eqref{eq:b-ell} such
that we arrive at
\begin{equation}
 r_\ell + \frac{2\kappa^{2\ell}}{\gamma^2} = \OO(\kappa) \,.
\end{equation}

\medskip
In all expressions above, the asymptotic normalization of course depends on
the binding momentum, \ie, $\gamma=\gamma(\kappa)$, only in general this
dependence can not be calculated analytically.  Using the results that we have
derived in this paper, it is possible to determine both $\kappa$ and $\gamma$
from the volume dependence of a given bound state.  If this state is
sufficiently shallow, the formulae derived in this chapter can then be used to
get an estimate for the effective range of the interaction.


\begin{thebibliography}{99}

\bibitem{Lee:2008fa}
  D.~Lee,
  Prog.\ Part.\ Nucl.\ Phys.\ {\bf 63} (2009) 117
  [arXiv:0804.3501 [nucl-th]].

\bibitem{Bazavov:2009bb}
  A.~Bazavov {\it et al.},
  Rev.\ Mod.\ Phys.\ {\bf 82} (2010) 1349
  [arXiv:0903.3598 [hep-lat]].

\bibitem{Beane:2010em}
  S.~R.~Beane, W.~Detmold, K.~Orginos and M.~J.~Savage,
  Prog.\ Part.\ Nucl.\ Phys.\ {\bf 66 } (2011) 1
  [arXiv:1004.2935 [hep-lat]].

\bibitem{Luscher:1985dn}
  M.~L{\"u}scher,
  Commun.\ Math.\ Phys.\ {\bf 104} (1986) 177.

\bibitem{Bour:2011ef}
  S.~Bour, S.~K{\"o}nig, D.~Lee, H.-W.~Hammer, and U.-G.~Mei{\ss}ner,
  [arXiv:1107.1272 [nucl-th]].

\bibitem{Davoudi:2011md}
  Z.~Davoudi and M.~J.~Savage,
  [arXiv:1108.5371 [hep-lat]].

\bibitem{Konig:2011nz}
  S.~K{\"o}nig, D.~Lee and H.-W.~Hammer,
  Phys.\ Rev.\ Lett.\ {\bf 107} (2011) 112001
  [arXiv:1103.4468 [hep-lat]].

\bibitem{Luscher:1991cf}
  M.~L\"uscher,
  Nucl.\ Phys.\ B\ {\bf 364} (1991) 237.
  
\bibitem{Bernard:2008ax}
  V.~Bernard, M.~Lage, U.-G.~Mei{\ss}ner and A.~Rusetsky,
  JHEP\ {\bf 0808} (2008) 024
  [arXiv:0806.4495 [hep-lat]].

\bibitem{Novikov:1977dq}
  V.~A.~Novikov \textit{et al.},
  Phys.\ Rept.\ {\bf 41} (1978) 1.

\bibitem{Bulava:2009jb}
  J.~Bulava {\it et al.},
  Phys.\ Rev.\ D\ {\bf 79} (2009) 034505
  [arXiv:0901.0027 [hep-lat]].

\bibitem{Matsuyama:2006rp}
  A.~Matsuyama, T.~Sato and T.-S.~H.~Lee,
  Phys.\ Rept.\ {\bf 439} (2007) 193
  [arXiv:nucl-th/0608051].

\bibitem{Beane:2011iw}
  S.~R.~Beane {\it et al.} [ NPLQCD Collaboration ],
  [arXiv:1109.2889 [hep-lat]].

\bibitem{Regal:2003zz}
  C.~A.~Regal, C.~Ticknor, J.~L.~Bohn and D.~S.~Jin,
  Phys.\ Rev.\ Lett.\ {\bf 90} (2003) 053201
  [arXiv:cond-mat/0209071].

\bibitem{Schunk:2005A}
  C.~H.~Schunck \textit{et al.},
  Phys.\ Rev.\ A {\bf 71} (2005) 045601
  [arXiv:cond-mat/0407373 [cond-mat.soft]].

\bibitem{Gaebler:2007A}
  J.~P.~Gaebler, J.~T.~Stewart, J.~L.~Bohn and D.~S.~Jin,
  Phys.\ Rev.\ Lett.\ {\bf 98} (2007) 200403
  [arXiv:cond-mat/0703087v2 [cond-mat.other]].

\bibitem{Riisager-94}
  K.~Riisager,
  Rev.\ Mod.\ Phys.\ {\bf 66} (1994) 1105.

\bibitem{Typel:2004zm}
  S.~Typel and G.~Baur,
  Phys.\ Rev.\ Lett.\ {\bf 93} (2004) 142502
  [arXiv:nucl-th/0406068].

\bibitem{Hammer:2011ye}
  H.-W.~Hammer and D.~R.~Phillips,
  Nucl.\ Phys.\ A\ {\bf 865} (2011) 17
  [arXiv:1103.1087 [nucl-th]].

\bibitem{Rupak:2011nk}
  G.~Rupak and R.~Higa,
  Phys.\ Rev.\ Lett.\ {\bf 106} (2011) 222501
  [arXiv:1101.0207 [nucl-th]].

\bibitem {Tohsaki:2001}
  A.~Tohsaki, H.~Horiuchi, P.~Schuck and G.~R{\"o}pke,
  Phys.\ Rev.\ Lett.\ {\bf 87} (2001) 192501
  [arXiv:nucl-th/0110014].

\bibitem {Chernykh:2007}
  M.~Chernykh \textit{et al.},
  Phys.\ Rev.\ Lett.\ {\bf 98} (2007) 032501.

\bibitem{Epelbaum:2011md}
  E.~Epelbaum, H.~Krebs, D.~Lee, U.-G.~Mei{\ss}ner,
  Phys.\ Rev.\ Lett.\ {\bf 106} (2011) 192501
  [arXiv:1101.2547 [nucl-th]].

\bibitem{Xu:1994zz}
  H.~M.~Xu, C.~A.~Gagliardi, R.~E.~Tribble, A.~M.~Mukhamedzhanov
  and N.~K.~Timofeyuk,
  Phys.\ Rev.\ Lett.\ {\bf 73} (1994) 2027.

\bibitem{Taylor}
  J.~R.~Taylor,
  {\it Scattering Theory: The Quantum Theory of Nonrelativistic Collisions},
  Dover Publications, Inc. (2006).

\bibitem{GbWatson}
  M.~L.~Goldberger and K.~M.~Watson,
  {\it Collision Theory},
  John Wiley \& Sons, Inc. (1967).

\bibitem{AbramStegPocket}
  M.~Abramowitz and I.~A.~Stegun,
  {\it Pocketbook of Mathematical Functions},
  Verlag Harri Deutsch, Thun; Frankfurt am Main (1984).

\bibitem{Luscher:1990ux}
  M.~L{\"u}scher,
  Nucl.\ Phys.\ B\ {\bf 354} (1991) 531.

\bibitem{Hammer:2010fw}
  H.-W.~Hammer and D.~Lee,
  Annals Phys.\ {\bf 325} (2010) 2212
  [arXiv:1002.4603 [nucl-th]].

\bibitem{Dyke:2011}
  P.~Dyke \textit{et al.},
  Phys.\ Rev.\ Lett.\ {\bf 106} (2011) 105304.

\bibitem{Frohlich:2011}
  B.~Fr{\"o}hlich \textit{et al.},
  Phys.\ Rev.\ Lett.\ {\bf 106} (2011) 105301.

\end{thebibliography}
\end{document}